\documentclass[12pt,nofootinbib,superscriptaddress]{revtex4}

\usepackage{amssymb,color,paralist,amsmath,epsfig}
\usepackage{graphicx}
\usepackage{amsfonts}
\usepackage{nicefrac,esint}
\usepackage{verbatim}
\usepackage{float}

\usepackage{relsize}

\usepackage[colorlinks,citecolor=blue,linktoc=all,linkcolor=cyan]{hyperref}
\usepackage[section]{placeins}

\newcommand{\cG}{{\cal G}}
\newcommand{\cF}{{\cal F}}

\newcommand{\beq}{\begin{equation}}
\newcommand{\eeq}{\end{equation}}

\newcommand{\ber}{\begin{eqnarray}} 
\newcommand{\eer}{\end{eqnarray}}

\usepackage{color}

\usepackage[normalem]{ulem}

\renewcommand\sout{\bgroup \color[rgb]{0,0.00,1.} \ULdepth=-.5ex \ULset}

\begin{document}

\title{Two-photon exchange contribution to elastic $e^-$-proton scattering: Full dispersive treatment of $\pi N$ states and comparison with data}
\author{Oleksandr Tomalak}
\affiliation{Institut f\"ur Kernphysik and PRISMA Cluster of Excellence, Johannes Gutenberg Universit\"at, Mainz, Germany}
\author{Barbara Pasquini}
\affiliation{Dipartimento di Fisica, Universit\`a degli Studi di Pavia, Pavia, Italy}
\affiliation{INFN Sezione di Pavia, Pavia, Italy}
\author{Marc Vanderhaeghen}
\affiliation{Institut f\"ur Kernphysik and PRISMA Cluster of Excellence, Johannes Gutenberg Universit\"at, Mainz, Germany}

\date{\today}

\begin{abstract}
We evaluate the two-photon exchange correction to the elastic electron-proton scattering cross section within a dispersive framework. Besides the elastic contribution, we account for all $\pi N$ intermediate state contributions using the phenomenological MAID fit as an input. We develop a novel method for the analytical continuation of the two-photon exchange amplitudes into the unphysical region and generalize our previous work to the momentum transfer region $ 0.064~\mathrm{GeV}^2 \lesssim Q^2 \lesssim 1~\mathrm{GeV}^2$. We compare our results with recent OLYMPUS, CLAS and VEPP-3 data as well as with empirical fits and estimates in the forward angular region.

\end{abstract}

\maketitle

\tableofcontents

\vspace{-0.2cm}

\section{Introduction}
\label{sec1}

The first measurements of the proton electromagnetic structure in terms of form factors (FFs) were performed by Hofstadter's group \cite{Mcallister:1956ng,Hofstadter:1956qs} using elastic scattering of electrons on protons under the assumption of the exchange of one virtual photon \cite{Rosenbluth:1950yq}. These experiments demonstrated that the proton has a finite size and allowed us to extract the Dirac FF $ F_D $ and Pauli FF $ F_P $ of the proton from cross section measurements at different electron scattering angles \cite{Hofstadter:1956qs}. This method has been refined over the years by many experiments. Currently, the most precise measurements of the proton FFs at low momentum transfer, and its charge and magnetic radii were performed by the A1 Collaboration at MAMI, Mainz \cite{Bernauer:2010wm,Bernauer:2013tpr}. The knowledge of the proton FFs reached a subpercent level accuracy yielding a proton charge radius from the electron-proton scattering: $ R_E = 0.879(8) ~\mathrm{fm}$ \cite{Bernauer:2010wm,Bernauer:2013tpr}. However, later reanalyses of the MAMI data gave different results in the range $ 0.84~\mathrm{fm} \lesssim R_E \lesssim 0.89~\mathrm{fm}$ \cite{Lorenz:2012tm,Lorenz:2014vha,Lorenz:2014yda,Lee:2015jqa,Arrington:2015ria,Arrington:2015yxa,Griffioen:2015hta,Higinbotham:2015rja}, mainly originating from the data extrapolation from the lowest accessible value of $ Q^2 = 5 \times 10^{-3}~\mathrm{GeV}^2$ down to $Q^2=0$. Besides this extrapolation issue, measurements of FFs with subpercent level accuracy raise the question of the theoretical control over corrections to the reaction formalism, notably radiative corrections. The leading radiative corrections which require a hadronic model for an estimate, and are thus not solely calculable in QED, are due to two-photon exchange (TPE) between the lepton and nucleon. \footnote{  The value of the two-photon exchange contribution depends on the applied radiative corrections and differs between the traditional Mo and Tsai \cite{Mo:1968cg} versus Maximon and Tjon \cite{Maximon:2000hm} prescriptions. 
In the soft-collinear effective field theory approach \cite{Hill:2016gdf}, 
a renormalization analysis was performed allowing us to systematically compute and resum large logarithms at momentum transfers 
$Q^2 \gg m_e^2$.}
The extraction of the charge radius, which relies mainly on forward angle elastic scattering data, does not show a significant model dependence for these TPE corrections. However, the value of the magnetic radius, which relies on backward scattering angle information, depends significantly on the applied TPE model \cite{Bernauer:2013tpr}. Besides this open question, the magnetic FF value extracted in Refs. \cite{Bernauer:2010wm,Bernauer:2013tpr} is systematically $2~\%$ larger for $Q^2 \gtrsim 0.2~\mathrm{GeV}^2$ when compared to results from previous measurements. These issues require us to reduce the model dependence in the treatment of TPE corrections to the elastic electron-proton scattering.

The recent extractions of the proton charge radius from the Lamb shift measurements in muonic hydrogen \cite{Pohl:2010zza,Antognini:1900ns} resulted in a significant discrepancy in comparison with measurements with electrons \cite{Bernauer:2010wm,Bernauer:2013tpr,Mohr:2012tt}, see Refs. \cite{Antognini:1900ns,Carlson:2015jba,Hill:2017wzi} for recent reviews. In view of this discrepancy, the higher-order corrections to the Lamb shift were examined in detail by many groups. In particular, the TPE proton structure correction was scrutinized over the past decade \cite{Pachucki:1996zza,Faustov:1999ga,Pineda:2002as,Pineda:2004mx,Nevado:2007dd,Carlson:2011zd,Hill:2012rh,Birse:2012eb,Alarcon:2013cba,Gorchtein:2013yga,Peset:2014jxa,Tomalak:2015hva,Caprini:2016wvy,Hill:2016bjv}. The TPE correction contributes at present the largest theoretical uncertainty when extracting the charge radius from the Lamb shift data, thus limiting its accuracy. However, its size is about ten times smaller than the observed discrepancy \cite{Carlson:2011zd}.
 
The precise knowledge of the elastic proton FFs, and its charge and magnetic radii is also of paramount importance in view of forthcoming high-precise measurements of the 1S hyperfine splitting by the CREMA Collaboration~\cite{Pohl:2016tqq}, FAMU Collaboration~\cite{Adamczak:2016pdb,Dupays:2003zz}, and a planned J-PARC experiment~\cite{Ma:2016etb}. These new experiments aim to measure the 1S hyperfine splitting to $1~\mathrm{ppm}$ accuracy largely exceeding the theoretical knowledge of the leading proton structure correction due to TPE \cite{Zemach:1956zz,Iddings:1959zz,Iddings:1965zz,Drell:1966kk,Faustov:1966,Faustov:1970,Bodwin:1987mj,Faustov:2001pn,Carlson:2008ke,Carlson:2011af,Hagelstein:2015egb,Peset:2016wjq,Tomalak:2017owk,Tomalak:2017npu}, which was estimated to be $213~\mathrm{ppm}$ in Ref. \cite{Carlson:2011af} and $102~\mathrm{ppm}$ in Ref. \cite{Tomalak:2017npu}. The uncertainty coming from the elastic proton structure of $48~\mathrm{ppm}$ \cite{Tomalak:2017npu} can be further reduced by new measurements of the electromagnetic FFs at low $Q^2$ \cite{Denig:2016dqo} and by reanalyzing the existing experimental data with improved treatment of TPE and higher-order QED radiative corrections. 

A second open question in the description of the proton electromagnetic structure arose in the beginning of this century after the realization that the polarization transfer from a longitudinally polarized electron to the proton, in the elastic scattering process, provided an alternative method to access the proton elastic FFs \cite{Akhiezer:1968ek,Akhiezer:1974em,Dombey:1969wk,Dombey:1969wi}. The ratio of the electric over magnetic FFs $G_{Ep} / G_{Mp}$ was measured at the Jefferson Lab in the scattering on the polarized proton and by the detection of the recoiling proton's polarization, see Ref.~\cite{Punjabi:2015bba} for a recent review. It was found that the measured ratio decreases approximately linearly with increasing momentum transfers~\cite{Jones:1999rz, Gayou:2001qd, Punjabi:2005wq, Puckett:2010ac} for $ Q^2 \gtrsim 1~\mathrm{GeV}^2$, in contradiction with the traditional extraction from unpolarized cross section measurements \cite{Rosenbluth:1950yq}, which shows an approximately constant behavior for the $G_{Ep}/G_{Mp}$ ratio. Apparently, the precise account of higher-order radiative corrections is necessary when going to larger momentum transfers. The unaccounted contribution from the hard two-photon exchange process was proposed as an explanation of this discrepancy~\cite{Guichon:2003qm, Blunden:2003sp}, which triggered a lot of research activity over the past years. 

In the relatively large momentum transfer region, the TPE correction was calculated theoretically \cite{Afanasev:2002gr,Blunden:2003sp,Gorchtein:2004ac,Pasquini:2004pv,Chen:2004tw, Afanasev:2005mp,Borisyuk:2008db,Kivel:2009eg,Kivel:2012vs} and studied experimentally \cite{Wells:2000rx,Maas:2004pd,Meziane:2010xc,Guttmann:2010au,BalaguerRios:2012uk,Abrahamyan:2012cg,Waidyawansa:2013yva,Kumar:2013yoa,Nuruzzaman:2015vba,Zhang:2015kna}, see Refs.~\cite{Carlson:2007sp, Arrington:2011dn} for reviews. Recently, three dedicated experiments confirmed the relevance of the TPE correction showing a deviation of the positron-proton to electron-proton elastic scattering cross section ratio from unity within $2\sigma$-$3\sigma$ (statistical and uncorrelated systematic errors). The measurements have been performed at VEPP-3~\cite{Rachek:2014fam},  by the CLAS Collaboration at JLab~\cite{Moteabbed:2013isu,Adikaram:2014ykv, Rimal:2016toz}, and by the OLYMPUS Collaboration at  DESY~\cite{Henderson:2016dea}, see Ref. \cite{Afanasev:2017gsk} for the most recent review of these experiments and a discussion of results.
 
At low momentum transfers, the leading term in the momentum transfer expansion of the TPE correction to the unpolarized electron-proton scattering cross section arises from the scattering of the relativistic massless electron on a point charged target \cite{McKinley:1948zz}, which is known as a Feshbach result. The subleading terms in the corresponding momentum transfer expansion of the TPE correction are due to finite size effects in the proton intermediate state, and due to inelastic intermediate states, see e.g. Ref. \cite{Brown:1970te}. Besides the leading term, proportional to $Q$, coming from the Feshbach correction, the expansion for the elastic (i.e. proton) intermediate state contains $Q^2 \ln^2 Q^2$ and $Q^2 \ln Q^2$ terms. The leading model-independent correction from all inelastic intermediate states is of order $ Q^2 \ln Q^2$, as first obtained in Ref.~\cite{Brown:1970te}, and subsequently reproduced within dispersion relations \cite{Gorchtein:2014hla}. Besides the leading inelastic corrections, the unpolarized proton structure function contribution, which enters at order $Q^2$, was evaluated in Refs.~\cite{Tomalak:2015aoa,Tomalak:2015hva}.

When going to larger momentum transfers, the TPE correction to the unpolarized elastic electron-proton scattering cross section was early on approximated as a nucleon box diagram with monopole FFs, which were evaluated using standard four-point integrals in Ref.~\cite{Blunden:2003sp}. This model was generalized to the case of the narrow-$\Delta$ intermediate state in Ref.~\cite{Kondratyuk:2005kk} with subsequent evaluations in Refs.~\cite{Graczyk:2013pca, Zhou:2014xka, Lorenz:2014yda}. Higher intermediate states were included in the work of Ref.~\cite{Kondratyuk:2007hc}, and a partial cancellation between the contributions from spin-1/2 and spin-3/2 resonances was found. However, the hadronic model calculations of Refs.~\cite{Blunden:2003sp,Kondratyuk:2005kk,Kondratyuk:2007hc,Graczyk:2013pca,Zhou:2014xka,Lorenz:2014yda} are based on the substitution of the off-shell vertex by its on-shell form unavoidably introducing model dependence. Such procedure can also result in pathological behavior as is e.g. the case for the TPE $ \Delta$-box contribution in the high-energy (HE) forward limit ($\varepsilon \to 1$) which diverges, violating unitarity \cite{Zhou:2014xka,Blunden:2017nby}.

The imaginary part of the TPE amplitudes can be obtained solely from the on-shell information by unitarity relations. Assuming the analyticity, the real part can then be reconstructed exploiting dispersion relations. Such approach for the TPE amplitudes was proposed in Refs.~\cite{Gorchtein:2006mq,Borisyuk:2008es}. The proton intermediate state (elastic) contribution was studied in Ref.~\cite{Borisyuk:2008es} and generalized to the case of spin-3/2 particles in Ref.~\cite{Borisyuk:2012he}. Higher spin-1/2 and spin-3/2 resonances were also accounted for in Refs.~\cite{Borisyuk:2013hja, Borisyuk:2015xma} exploiting the empirical multipoles for pion electroproduction. In the developed approach of Refs. \cite{Borisyuk:2008es,Borisyuk:2012he,Borisyuk:2013hja,Borisyuk:2015xma}, the experimental input was reparametrized as a sum of monopole FFs reducing the calculation to the evaluation of one-loop box diagrams as it is done in the hadronic models.

The data-driven dispersion relation approach, aimed at evaluating the dispersive integral directly from the experimental input, was presented in Ref.~\cite{Tomalak:2014sva} for the elastic intermediate state TPE contribution and generalized to the case of the narrow-$\Delta$ TPE in Ref.~\cite{Blunden:2017nby}. Within a dispersive framework one requires also the knowledge of the imaginary part of the TPE amplitudes outside the physical region for $ e p \to e p$ scattering. For one-particle intermediate states, the method of the analytical continuation of the TPE amplitudes into the unphysical region was described in Refs. \cite{Tomalak:2014sva,Blunden:2017nby}. 

\begin{figure}[H]
\begin{center}
\includegraphics[width=.44\textwidth]{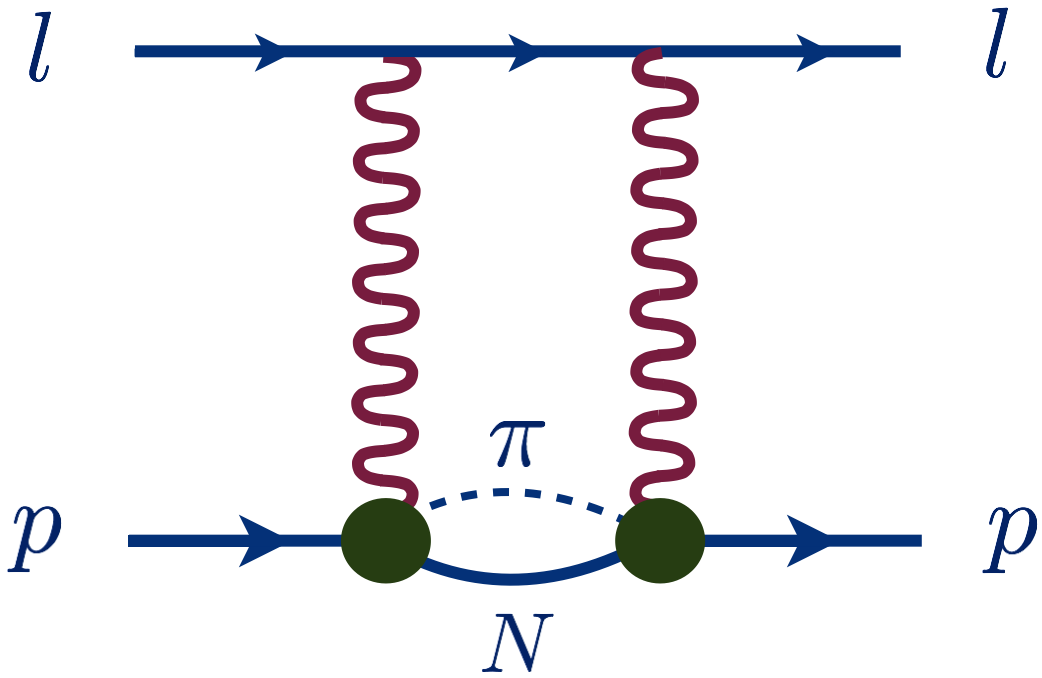}
\end{center}
\caption{TPE graph with $ \pi N $ intermediate state.}
\label{piN_TPE}
\end{figure}

A first step to extend such dispersive approach beyond narrow resonances was performed in Ref.~\cite{Tomalak:2016vbf}. In that work, the full $\pi N$ intermediate state TPE contribution, see Fig. \ref{piN_TPE}, was evaluated at low momentum transfer $ Q^2 \lesssim 0.064~\mathrm{GeV}^2$, where the analytical continuation of TPE amplitudes into the unphysical region is not required. Such approach allows us to account for all known $\pi N$ resonances with spins 1/2, 3/2, 5/2, ... as well as nonresonant $\pi N$ states. The pion electroproduction amplitudes from the MAID2007 fit~\cite{Drechsel:1998hk,Drechsel:2007if} were used as input in Ref. \cite{Tomalak:2016vbf} to evaluate the imaginary parts of the TPE amplitudes~\cite{Pasquini:2004pv}. It was found that the account for a $\pi N$ intermediate state at low momentum transfer within the subtracted DRs improves the agreement with fits to the experimental data \cite{Tomalak:2016vbf}.

In the present work, we extend the dispersion relation formalism of Ref.~\cite{Tomalak:2016vbf} to the momentum transfer range $ 0.064~\mathrm{GeV}^2 \lesssim Q^2 \lesssim 1~\mathrm{GeV}^2$. As a necessary step, we develop and test a novel method for the analytical continuation of the TPE amplitudes, which allows us to reconstruct the imaginary parts, exploiting the most recent pion electroproduction amplitudes from the MAID2007 fit~\cite{Drechsel:1998hk,Drechsel:2007if} as input, without having to approximate the resonance production FFs by sums of monopoles. Subsequently, we compare the sum of elastic and $\pi N$ intermediate state TPE corrections to recent experimental data as well as to the total TPE contribution in the near-forward approximation of Ref.~\cite{Tomalak:2015aoa}.

The paper is organized as follows: The general formalism of the elastic electron-proton scattering and of TPE corrections to observables are described in Sec. \ref{sec2}. The model calculation of the $\Delta$(1232) resonance contribution is given in Sec. \ref{sec3}. We study the $\Delta$(1232) resonance in the simplified hadronic model in Sec. \ref{sec31}, compare it with the unitarity relations in Sec. \ref{sec32} and to the dispersion relation approach in Sec. \ref{sec33}. Using this model calculation as a test case, we develop a novel method for the analytical continuation of the TPE amplitudes into the unphysical region in Sec. \ref{sec34}. In the following Sec. \ref{sec4}, we apply this method to evaluate the $ \pi N $ contribution, using the phenomenological $\pi N$ electroproduction multipoles from the MAID2007 fit as input, to the imaginary parts of TPE amplitudes. We determine the corresponding TPE corrections to observables. We compare our results with recent OLYMPUS, CLAS and VEPP-3 data and with polarization transfer measurements in Sec. \ref{sec5}. We also provide a comparison with the empirical fits of Refs. \cite{Bernauer:2010wm,Bernauer:2013tpr} and total TPE calculation in the forward angular region of Ref. \cite{Tomalak:2015aoa}. We provide our conclusions and outlook in Sec. \ref{sec6}.

\section{Elastic $ ep$ scattering and TPE correction}
\label{sec2}

\begin{figure}[h]
\begin{center}
\includegraphics[width=.45\textwidth]{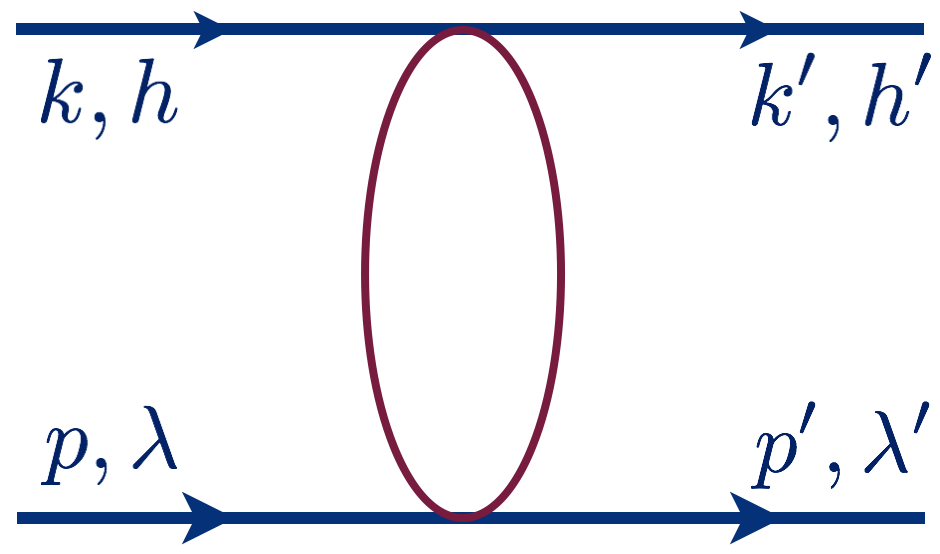}
\end{center}
\caption{Elastic electron-proton scattering.}
\label{elastic_scattering_general}
\end{figure}

The elastic electron-proton scattering process: $ e( k , h ) + p( p, \lambda ) \to e( k^\prime, h^\prime) + p(p^\prime, \lambda^\prime) $, where $ k,~p,~k^\prime,~p^\prime$ denote the participating particles momenta, $ h(h^\prime) $ the incoming (outgoing) electron helicities and $ \lambda(\lambda^\prime) $ the corresponding proton helicities respectively, see Fig. \ref{elastic_scattering_general}, is completely described by 2 Mandelstam variables. Conveniently, we work with the squared momentum transfer $ Q^2 = - (k-k^\prime)^2$ and  the squared energy in the center-of-mass (c.m.) reference frame $ s = ( p + k )^2 $.

The symmetry between the $s$ and $u$ channels can easily be incorporated introducing the crossing-symmetric kinematical variable $ \nu$:
\ber
 \nu \equiv ( s - u )/4,
\eer
where $ u = ( k - p^\prime )^2 $. In the experimental analyses, it is convenient to introduce the photon polarization parameter $ \varepsilon$, which indicates the degree of the longitudinal polarization of the virtual photon:
\ber
\varepsilon & = & \frac{ \nu^2 - M^4 \tau_P ( 1 + \tau_P )}{ \nu^2 + M^4 \tau_P ( 1 + \tau_P )} ,
\eer
with the proton mass $M$, and $\tau_P\equiv Q^2/(4M^2)$. It varies between $\varepsilon = 0$ for backward scattering and $\varepsilon = 1$ for forward scattering.

The elastic $ e^{-} p $ scattering with massless leptons is completely described by three independent Lorentz-invariant 
amplitudes~\cite{Guichon:2003qm}:
\begin{eqnarray}
 \label{str_ampl} 
T  &=&  
\frac{e^2}{Q^2} \bar{u}(k^\prime,h^\prime) \gamma_\mu u(k,h) \nonumber\\
&\times & \bar{N}(p^\prime,\lambda^\prime) 
 \left( \gamma^\mu \cG_M (\nu, Q^2) -   \frac{P^{\mu}}{M} \cF_2 (\nu, Q^2) 
 +  \frac{\gamma. K P^{\mu}}{M^2} 
  \cF_3 (\nu, Q^2) \right)N(p,\lambda),
  \label{str_ampl1}
\end{eqnarray}
where the averaged momentum variables are $ P = (p+p^\prime)/2, ~ K = (k+k^\prime)/2 $; $u$ ($ \bar{u}$) is the initial (final) electron spinor; $N$ ($ \bar{N}$) is the initial (final) proton spinor; $\gamma. a  \equiv \gamma^\mu a_\mu$; and $e > 0$ is the proton charge. In the following, we consider also the amplitudes $ \cG_1$ and $ \cG_2 $, defined by
\ber 
 {\cal{G}}_{1}  & \equiv & {\cal{G}}_{M}  + \frac{\nu}{M^2} {\cal{F}}_3,  \label{new_amplitude1}  \\ 
 {\cal{G}}_{2}  & \equiv & {\cal{G}}_{M}  - \left( 1 + \tau_P \right) {\cal{F}}_2  + \frac{\nu}{M^2} {\cal{F}}_3. \label{new_amplitude2}
\eer
In the approximation of one-photon exchange (OPE), these amplitudes are equivalent to the magnetic $ \cG^{1 \gamma}_1  = \cG_M^{1 \gamma}  = G_M (Q^2) $ and electric $ \cG^{1 \gamma}_2 = G_E (Q^2) \equiv G_M ( Q^2 )- ( 1 + \tau_P) F_2 (Q^2) $ proton FFs, where $ F_2 (Q^2) = \cF_2^{1 \gamma} $ is the Pauli FF. The amplitude $ \cF_3 $ vanishes in the OPE approximation: $ \cF^{1 \gamma}_3 = 0 $.

In presence of TPE, the $e^- p \to e^- p$ elastic scattering cross section receives corrections which can be expressed as
\ber
\sigma = \sigma_\mathrm{OPE} \left( 1 + \delta_{2 \gamma} \right),
\eer
where $\sigma_\mathrm{OPE}$ is the cross section in the OPE approximation. In terms of the invariant amplitudes, the TPE correction $\delta_{2 \gamma}$ to the unpolarized $ e^- p $ cross section at the leading order in $ \alpha  \equiv e^2 / 4 \pi \simeq 1/137 $ is given by \cite{Tomalak:2016vbf}
\beq \label{unpolarized_cross_section}
\delta_{2 \gamma} = \frac{2}{G^2_M  + \frac{\varepsilon}{\tau_P} G^2_E } \left\{ G_M  \Re  {\cal{G}}_{1}^{2 \gamma}  + \frac{\varepsilon}{\tau_P} {G_E}  { \Re  \cal{G}}_{2}^{2 \gamma}  +  {G_M} \left (\varepsilon -1\right) \frac{\nu}{M^2} {{\Re \cal{F}}_3^{2 \gamma}}  \right\},
\eeq
where the superscript $2 \gamma$ on the invariant amplitudes indicates their TPE contributions.

Other accessible observables, which are influenced by the real parts of the TPE amplitudes, are double polarization observables with a polarization transfer from the longitudinally polarized electron to the recoil proton. The longitudinal polarization transfer asymmetry is defined as
\ber
P_l = \frac{\mathrm{d} \sigma \left( h = +,~\lambda'=+ \right) - \mathrm{d} \sigma \left( h = +,~\lambda'=- \right)}{\mathrm{d} \sigma \left( h = +,~\lambda'=+ \right) + \mathrm{d} \sigma \left( h = +,~\lambda'=- \right)},
\eer
and the transverse polarization transfer asymmetry is given by
\ber
P_t = \frac{\mathrm{d} \sigma \left( h = +,~S'=S_\perp \right) - \mathrm{d} \sigma \left( h = +,~S'=-S_\perp \right)}{\mathrm{d} \sigma \left( h = +,~S'=S_\perp \right) + \mathrm{d} \sigma \left( h = +,~S'= - S_\perp \right)},
\eer
with the spin direction of the recoil proton $ S' = \pm S_\perp $ in the scattering plane transverse to its momentum direction.

In this work, we also discuss the ratio of polarization transfer asymmetries, which is measured experimentally \cite{Meziane:2010xc}:
\ber \label{polarization_observables1}
 - \sqrt{\frac{\tau_P ( 1 + \varepsilon )}{2 \varepsilon}} \frac{P_t}{P_l} & = &  \frac{G_E}{G_M} + \frac{ \Re \cG_2^{2 \gamma}  }{G_M} -\frac{G_E}{G_M} \frac{ \Re \cG_1^{2 \gamma}  }{G_M} + \frac{ 1 - \varepsilon }{ 1 + \varepsilon } \frac{G_E}{G_M} \frac{\nu}{M^2} \frac{ \Re \cF_3^{2 \gamma}  }{G_M}.
\eer

\section{$\Delta$(1232) contribution}
\label{sec3}

In this section, we study the prominent $\Delta$(1232) resonance contribution to TPE amplitudes for elastic electron-proton scattering. First, we describe a model calculation of the narrow-$\Delta$ TPE correction \cite{Kondratyuk:2005kk}. This model will firstly serve the purpose to provide a detailed comparison with the dispersion relation (DR) approach \cite{Borisyuk:2012he,Blunden:2017nby}. Afterwards, we develop a new method for the analytical continuation of the imaginary parts of the TPE amplitudes outside the physical region for the $e p \to e p$ scattering process. We will test this new method on the example of the $\Delta$ resonance contribution, where we know the amplitudes both in the physical and unphysical regions from the direct loop calculation. Once this method has been tested on the $\Delta$-intermediate state, we can be confident to apply it for the $\pi N$ intermediate state contribution to the TPE amplitudes in Sec. \ref{sec4}.

\subsection{Box graph model}
\label{sec31}

\begin{figure}[H]
\begin{center}
\includegraphics[width=.95\textwidth]{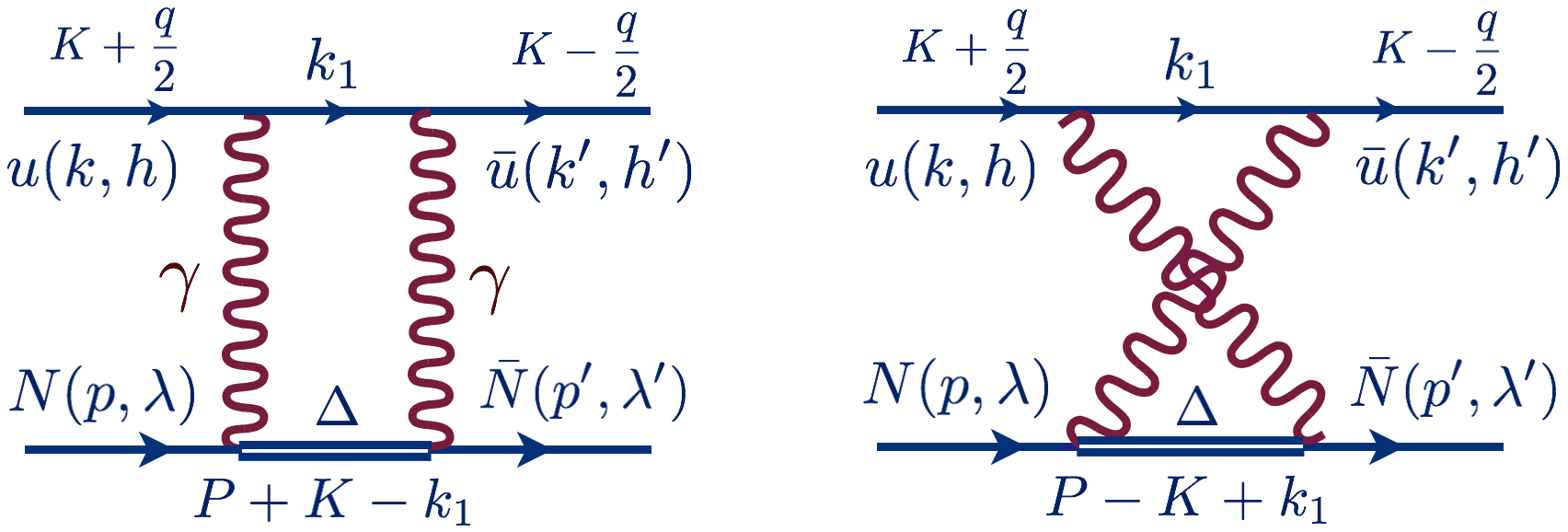}
\end{center}
\caption{Direct and crossed TPE diagrams with $\Delta $-intermediate state in the elastic $ e p $ scattering.}
\label{model_graph}
\end{figure}

In this section, we use a box graph model to evaluate the narrow-$\Delta$ contribution to the TPE correction in the elastic electron-proton scattering at low momentum transfer, see Fig. \ref{model_graph}.

To model the $ \gamma^* N \to \Delta $ vertex, we restrict ourselves to the leading magnetic dipole transition,
\ber \label{OPE_Delta}
\Gamma^{ \alpha \mu}_{N \Delta} = \sqrt{\frac{2}{3}} \frac{3 \left( M+ M_\Delta \right)  {G}^*_\mathrm{M} \left( Q^2 \right) }{ 2 M \left(\left( M+ M_\Delta \right)^2 + Q^2 \right)}  \varepsilon^{\alpha \mu \rho \sigma} \left( p_\Delta \right)_\rho \tilde{q}_\sigma, \qquad \tilde{q} = p_\Delta - p,
\eer
using the on-shell magnetic transition FF $ G^*_\mathrm{M} \left( Q^2 \right) $, in the Jones-Scadron convention \cite{Jones:1972ky}, where $M_\Delta$ is the $ \Delta$ mass.

In this model, the helicity amplitudes corresponding with the TPE direct and crossed box graphs can be expressed as
\ber
 T^{2\gamma}_{\mathrm{direct}}  & =  & - i e^4 \mathop{\mathlarger{\int}}  \frac{\mathrm{d}^4 k_1}{( 2 \pi )^4} \bar{u}(k^\prime,h^\prime) \gamma_\nu \frac{\gamma.k_1}{k_1^2 - m^2} \gamma_\mu u (k,h)\frac{1}{(k_1 - K - \frac{q}{2} )^2(k_1 - K + \frac{q}{2} )^2 }\nonumber \\
& \times &  \bar{N}(p^\prime,\lambda^\prime) \left(\gamma^0 \Gamma^{\beta \nu}_{N \Delta} \gamma^0 \right)^\dagger \frac{\gamma.P + \gamma.K - \gamma.k_1 + M_\Delta}{(P + K - k_1)^2 - M_\Delta^2}  \left(  - g_{\beta \alpha} + \frac{1}{3} \gamma_\beta \gamma_\alpha \right)  \Gamma^{\alpha \mu}_{N \Delta} N (p,\lambda),  \nonumber \\ \label{helamp} \\
 T^{2\gamma}_{\mathrm{crossed}}  & = & - i e^4 \mathop{\mathlarger{\int}}  \frac{\mathrm{d}^4 k_1}{( 2 \pi )^4} \bar{u}(k^\prime,h^\prime) \gamma_\mu \frac{\gamma.k_1}{k_1^2 - m^2} \gamma_\nu u (k,h) \frac{1}{(k_1 - K - \frac{q}{2} )^2(k_1 - K + \frac{q}{2} )^2 } \nonumber \\
& \times & \bar{N} (p^\prime,\lambda^\prime) \left(\gamma^0 \Gamma^{\beta \nu}_{N \Delta} \gamma^0 \right)^\dagger \frac{\gamma.P - \gamma.K + \gamma.k_1 + M_\Delta}{(P - K + k_1)^2 - M_\Delta^2}  \left(  - g_{\beta \alpha} + \frac{1}{3} \gamma_\beta \gamma_\alpha \right)   \Gamma^{\alpha \mu}_{N \Delta} N (p,\lambda), \nonumber \\ \label{helamp1}
\eer
where $ P $ and $ K $ are defined as in Sec. \ref{sec2} and $m$ denotes the mass of the electron. In Eqs. \eqref{helamp} and \eqref{helamp1}, the simplified form of the vertex made it possible to replace the projection operator on the spin-3/2 states in the $\Delta$ propagator by
\ber
 \Pi_{\beta \alpha} (p_\Delta) = - g_{\beta \alpha} + \frac{1}{3} \gamma_\beta \gamma_\alpha  + \frac{\hat{p}_\Delta \gamma_\beta (p_\Delta)_\alpha + (p_\Delta)_\beta \gamma_\alpha \hat{p}_\Delta}{3 p_\Delta^2} \to - g_{\beta \alpha} + \frac{1}{3} \gamma_\beta \gamma_\alpha.
\eer

We next evaluate the TPE invariant amplitudes from the helicity amplitudes of Eqs. (\ref{helamp}), (\ref{helamp1}) \cite{Tomalak:2016vbf}. In the $ \gamma^* N \to \Delta $ vertex of Eq. (\ref{OPE_Delta}), the magnetic transition Jones and Scadron FF $ G^*_\mathrm{M} \left( Q^2 \right) $ is expressed in terms of the proton and neutron elastic Pauli FFs $ F_2^p$ and $ F_2^n$, respectively, using a large-$N_c$ theory relation \cite{Pascalutsa:2007wz}:
\ber \label{real_delta}
G^*_\mathrm{M} \left( Q^2 \right) & = & \frac{G^*_\mathrm{M} \left( 0 \right) }{\mu_p - \mu_n - 1} \left( F^p_2 (Q^2) - F^n_2 (Q^2) \right), \qquad G^*_\mathrm{M} \left( 0 \right) = 3.02,  \label{magnetic_transition_FF} \\
F^p_2 (Q^2) & = & \frac{\mu_p - 1}{\left( 1 +\tau_P \right) \left( 1 + \frac{Q^2}{\Lambda^2} \right)^2}, \qquad \mu_p = 2.793, \qquad \Lambda = 0.843~\mathrm{GeV}, \\
F^n_2 (Q^2) & = & \frac{\mu_n}{\left( 1 + \tau_P \right) \left( 1 + \frac{Q^2}{\Lambda^2} \right)^2} \left( \frac{ 1 + \left( a + b \right) \tau_P}{ 1 + b \tau_P } \right) , \quad \mu_n = -1.913, \quad a = 1.25, \quad b = 18.3, \nonumber \\
\eer
where the neutron electric FF is taken from Ref.~\cite{Warren:2003ma}. For the neutron magnetic, proton electric, as well as proton magnetic FFs a dipole form is assumed.

To calculate the invariant amplitudes, we use the four-point integrals from LoopTools \cite{Hahn:2000jm,vanOldenborgh:1989wn}. We confirm that the box graph with $ \Delta $-intermediate state is free of infrared divergencies, as is expected. We checked numerically that the amplitudes $ \cG^{2\gamma}_1,~\cG^{2\gamma}_2,~\cF^{2\gamma}_2 $ vanish in the limit $ Q^2 \to 0 $ at a fixed value of $\nu$, whereas the amplitudes $ \cG^{2\gamma}_M,~\cF^{2\gamma}_3 $ behave as $ a \ln Q^2 + b $, where $a$ and $b$ are constants, in agreement with the low-$ Q^2 $ limit of Ref.~\cite{Tomalak_PhD} and the results reported in Ref.~\cite{Tomalak:2016vbf}. In the following sections, we compare this model calculation, with one-loop integrals evaluated using LoopTools, to the dispersion relation evaluation.

\subsection{Unitarity relations}
\label{sec32}

In this section, we check numerically that the imaginary parts of the TPE amplitudes in the box graph model of Sec. \ref{sec31} are reconstructed by unitarity relations. We also compare the narrow-$\Delta$ model with a weighted-$\Delta$ model, as well as with the leading pion electroproduction multipole $ M^{(3/2)}_{1+} $ contribution, obtained from data.

To write down the unitarity relations directly for $ \Im T^{2 \gamma}_{\mathrm{direct}}$ from Eq. (\ref{helamp}), we exploit Cutkosky's rules by putting the intermediate state on its mass shell, i.e. replacing the intermediate electron and $\Delta$ propagator denominators in the loop integral by
\ber
\frac{1}{k_1^2 - m^2} & \to & ( - 2 \pi i ) \delta (k_1^2 - m^2) \Theta(k_1^0), \\
\frac{1}{(P + K - k_1)^2 - M_\Delta^2} & \to &  ( - 2 \pi i ) \delta ((K+P-k_1)^2 - M_\Delta^2) \Theta(\sqrt{s} - k_1^0).
\eer

Performing the integration over the electron energy and absolute value of the momentum, we obtain for the imaginary part of the TPE amplitude $\Im T^{2\gamma}_{\mathrm{direct}}$:
\ber
\Im T^{2\gamma}_{\mathrm{direct}}  & =  & \frac{\alpha^2 \left( s-M^2_\Delta \right)}{4 s} \mathop{\mathlarger{\int}}  \frac{\mathrm{d} \Omega_1}{(k_1 - K - \frac{q}{2} )^2(k_1 - K + \frac{q}{2} )^2 } \bar{u}(k^\prime,h^\prime) \gamma_\nu \gamma.k_1 \gamma_\mu u (k,h)\nonumber \\
& \times &  \bar{N}(p^\prime,\lambda^\prime) \left(\gamma^0 \Gamma^{\beta \nu}_{N \Delta} \gamma^0 \right)^\dagger \left( \gamma.P + \gamma.K - \gamma.k_1 + M_\Delta \right)  \left(  - g_{\beta \alpha} + \frac{1}{3} \gamma_\beta \gamma_\alpha \right)  \Gamma^{\alpha \mu}_{N \Delta} N (p,\lambda), \nonumber \\ \label{unitarity0}
\eer
where the integration runs over the intermediate electron angles $ \Omega_1$.

\begin{figure}[ht]
\begin{center}
\includegraphics[width=.48\textwidth]{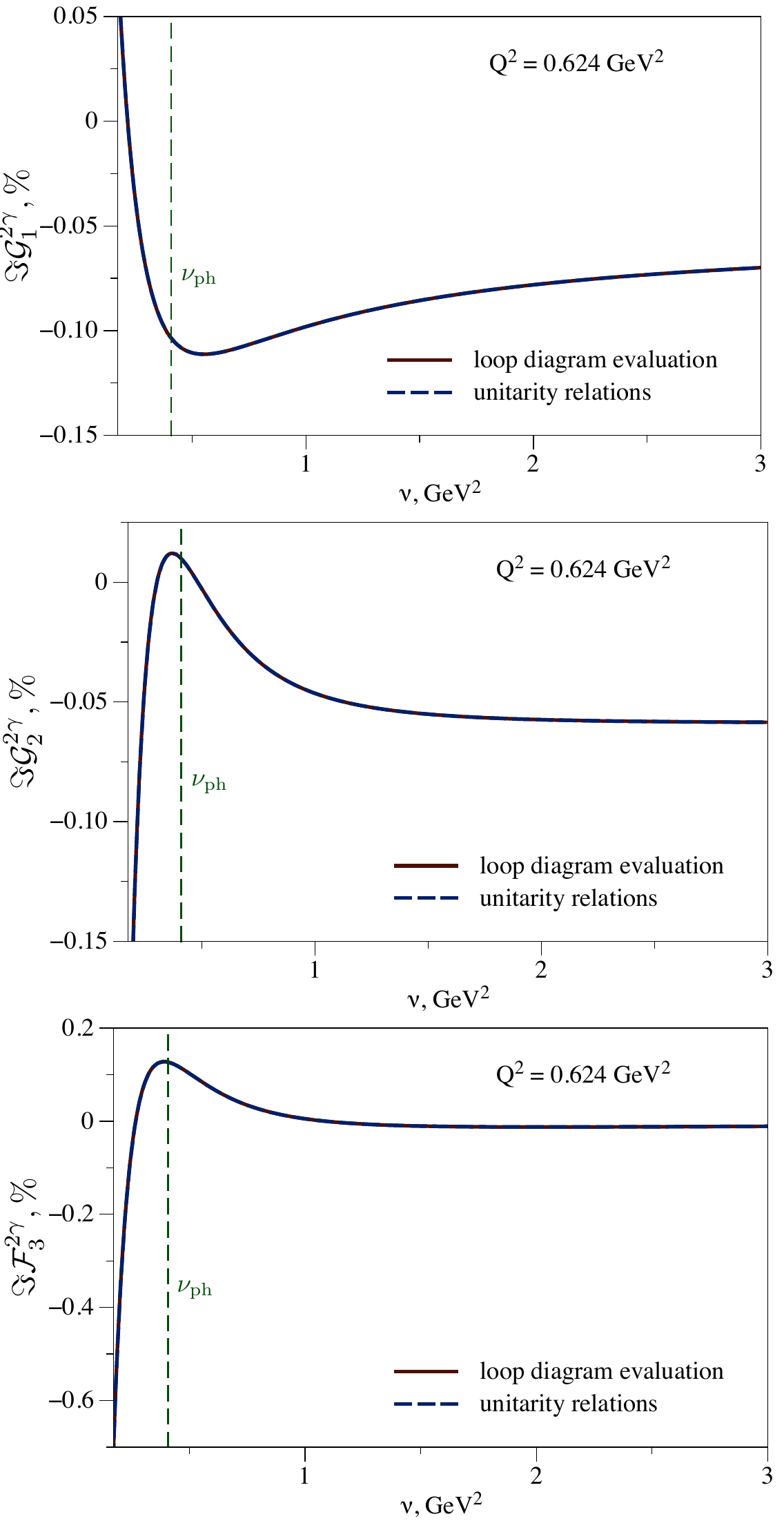}
\end{center}
\caption{Imaginary parts of the narrow-$\Delta$ contribution to the TPE amplitudes as a function of the crossing-symmetric variable $\nu$ in the physical and unphysical regions. The result of the direct box graph model evaluation is compared with the result obtained from the unitarity relations for $ Q^2 = 0.624~\mathrm{GeV}^2$. The vertical line corresponds with the boundary between physical ($\nu > \nu_\mathrm{ph}$) and unphysical ($\nu < \nu_\mathrm{ph}$) regions, where $ \nu_{\mathrm{ph}} = \sqrt{Q^2 \left( Q^2 + 4 M^2 \right)}/4 \approx 0.402 ~\mathrm{GeV}^2 $.}
\label{imaginary_comparison_Q2_0_624}
\end{figure}

We checked explicitly that the imaginary parts of the TPE amplitudes in the direct loop diagram evaluation within the box graph model of Sec. \ref{sec31} are in agreement with the unitarity relations of Eq. (\ref{unitarity0}) in the physical region, i.e., when the kinematics correspond to the geometrically allowed configuration for the $e p \to e p$ process. Performing the analytical continuation into the unphysical region by the contour deformation method discussed in Refs.~\cite{Tomalak:2014sva,Blunden:2017nby}, we evaluated the imaginary parts of the TPE amplitudes for arbitrary values of the crossing-symmetric variable $\nu$ based on the unitarity relation of Eq. (\ref{unitarity0}). We refer the reader to Refs.~\cite{Tomalak:2014sva,Tomalak_PhD,Tomalak:2016vbf} for a detailed description of the unphysical region for a narrow hadronic intermediate state. In Fig. \ref{imaginary_comparison_Q2_0_624} we show the comparison for the imaginary part calculated from Eq. (\ref{unitarity0}) through contour deformation with the direct loop diagram evaluation from the box graph model expression of Eq. (\ref{helamp}) for a narrow-$\Delta$ state using LoopTools \cite{Hahn:2000jm,vanOldenborgh:1989wn}. Using as an example the value of the momentum transfer $ Q^2 = 0.624~\mathrm{GeV}^2$, corresponding with 
a kinematics of the OLYMPUS experiment, we find a perfect agreement between both calculations.

\begin{figure}[H]
\begin{center}
\includegraphics[width=0.9\textwidth]{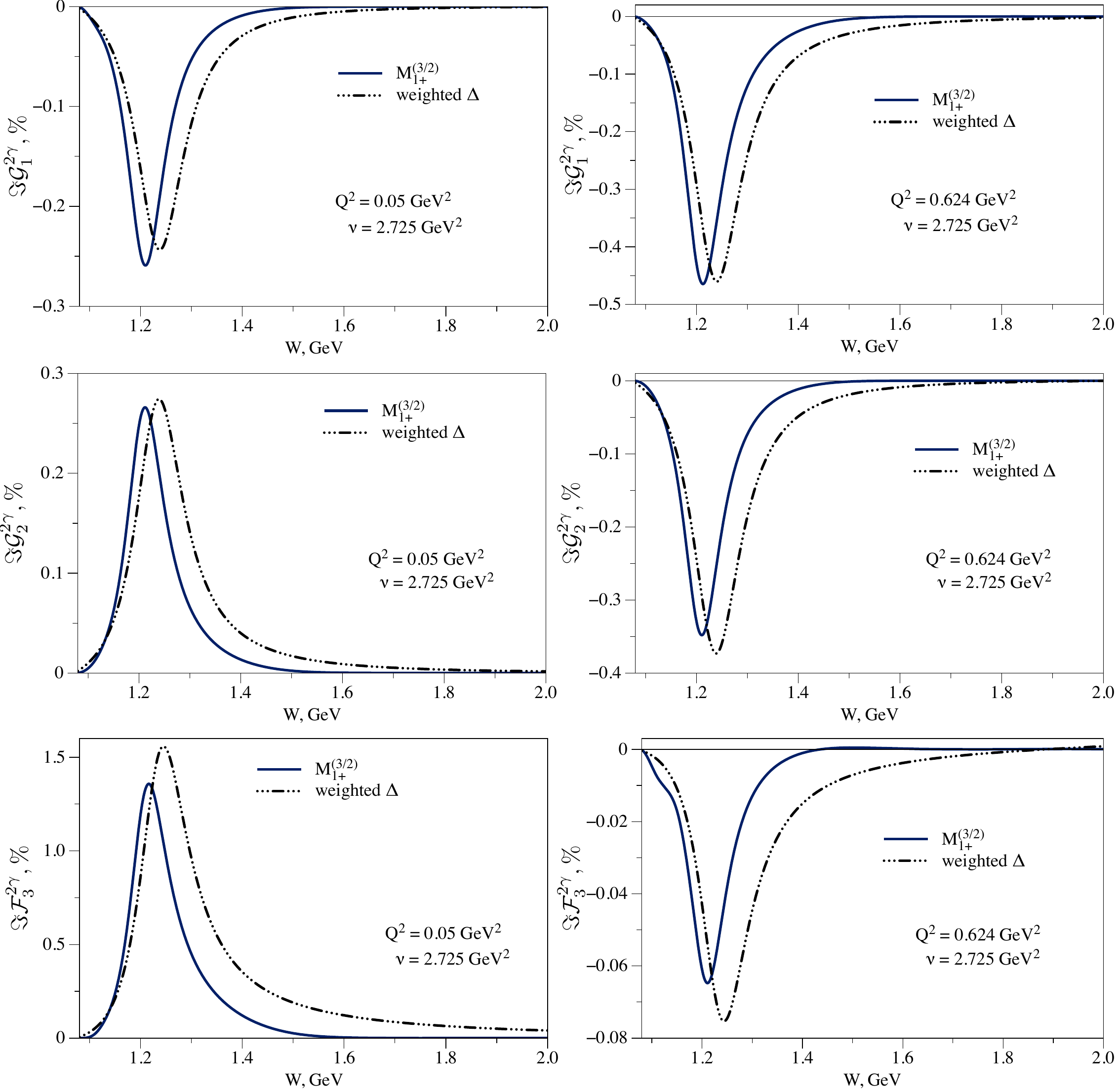}
\end{center}
\caption{$W$ integrand entering the imaginary parts of the TPE amplitudes for $\nu = 2.725~\mathrm{GeV}^2$ and for two $Q^2$ values: $ Q^2 = 0.05~\mathrm{GeV^2}$ (left panel) and $ Q^2 = 0.624~\mathrm{GeV^2}$ (right panel). The weighted-$\Delta$ result is compared with the full $ M^{(3/2)}_{1+}$ $ \pi N$ multipole contribution, as calculated from the MAID 2007 fit~\cite{Drechsel:1998hk,Drechsel:2007if}, which also includes nonresonant contributions.}
\label{imaginary_weight}
\end{figure}

In order to compare the $\Delta$ calculation with the empirical $\pi N$ multipole evaluation, we also consider in the following the more realistic $ \Delta $ contribution weighted over the invariant mass of the intermediate hadronic state: $ W^2 = (P+K-k_1)^2 $. In this case, the phase space integral entering the $ \Im T^{2 \gamma}_\mathrm{direct}$ in Eq. (\ref{unitarity0}) gets replaced by
\ber
\int \mathrm{d} \Omega_1 ... \to \int \limits_{M + m_\pi}^{\sqrt{s}}  \mathrm{d} W f \left( W \right) \int \mathrm{d} \Omega_1 ...,
\eer
where the weighting function $ f(W)$ is given by the Breit-Wigner form:
\ber \label{narrow_delta_weight}
f \left( W \right) & = & \frac{N_\Delta  }{W^6} \frac{\left( W^2 - M^2 + m^2_\pi \right)^2 - 4 W^2 m^2_\pi}{\left( W^2 - M_\Delta^2 \right)^2 + M^2_\Delta \Gamma_\Delta^2} \Theta \left( W - M - m_\pi \right).
\eer
Furthermore, we use as parameter values the pion mass $ m_\pi \approx 0.135~\mathrm{GeV}$; the $\Delta$ mass $ M_\Delta = 1.232~\mathrm{GeV}$, the $\Delta$ width $\Gamma_\Delta = 0.117~\mathrm{GeV}$; and the normalization parameter $ N_\Delta = \left( \int \limits_{M+m_\pi}^{\infty} f \left(W\right) \mathrm{d} W \right)^{-1}$. The weighting function of Eq. (\ref{narrow_delta_weight}) inherits the correct resonance shape and width as well as the correct behavior near the pion-production threshold $ W = M + m_\pi$. We adopt an overall prefactor $W^{-6}$ in order to have a comparable strength at the peak position as the $ M^{(3/2)}_{1+}$ $ \pi N$ contribution \cite{Tomalak:2016vbf}, which is evaluated with the MAID2007 fit~\cite{Drechsel:1998hk,Drechsel:2007if} as an input. We show this comparison in Fig. \ref{imaginary_weight}, where the $W$ distribution of the TPE amplitudes is presented for $\nu = 2.725~\mathrm{GeV}^2$ and $ Q^2 = 0.05~\mathrm{GeV^2}$ for the weighted-$\Delta$ model and for the $ M^{(3/2)}_{1+}$ $ \pi N$ contribution. Both calculations show approximately the same strength at the $\Delta$-resonance position. The  shift of the peak position in the empirical MAID fit is understood to be due to the unitarization between the resonant contribution and nonresonant background, which are both present in the $ M^{(3/2)}_{1+}$ multipole.

In the following Fig. \ref{imaginary_comparison_Q2_0_624_weight}, we compare the imaginary parts of TPE amplitudes as calculated using the narrow-$\Delta $ model, the weighted-$\Delta$ model, and using the dominant magnetic dipole $ M^{(3/2)}_{1+}$ $ \pi N$ contribution \cite{Tomalak:2016vbf}, which is evaluated from the MAID2007 fit~\cite{Drechsel:1998hk,Drechsel:2007if} as an input. 

\begin{figure}[H]
\begin{center}
\includegraphics[width=.45\textwidth]{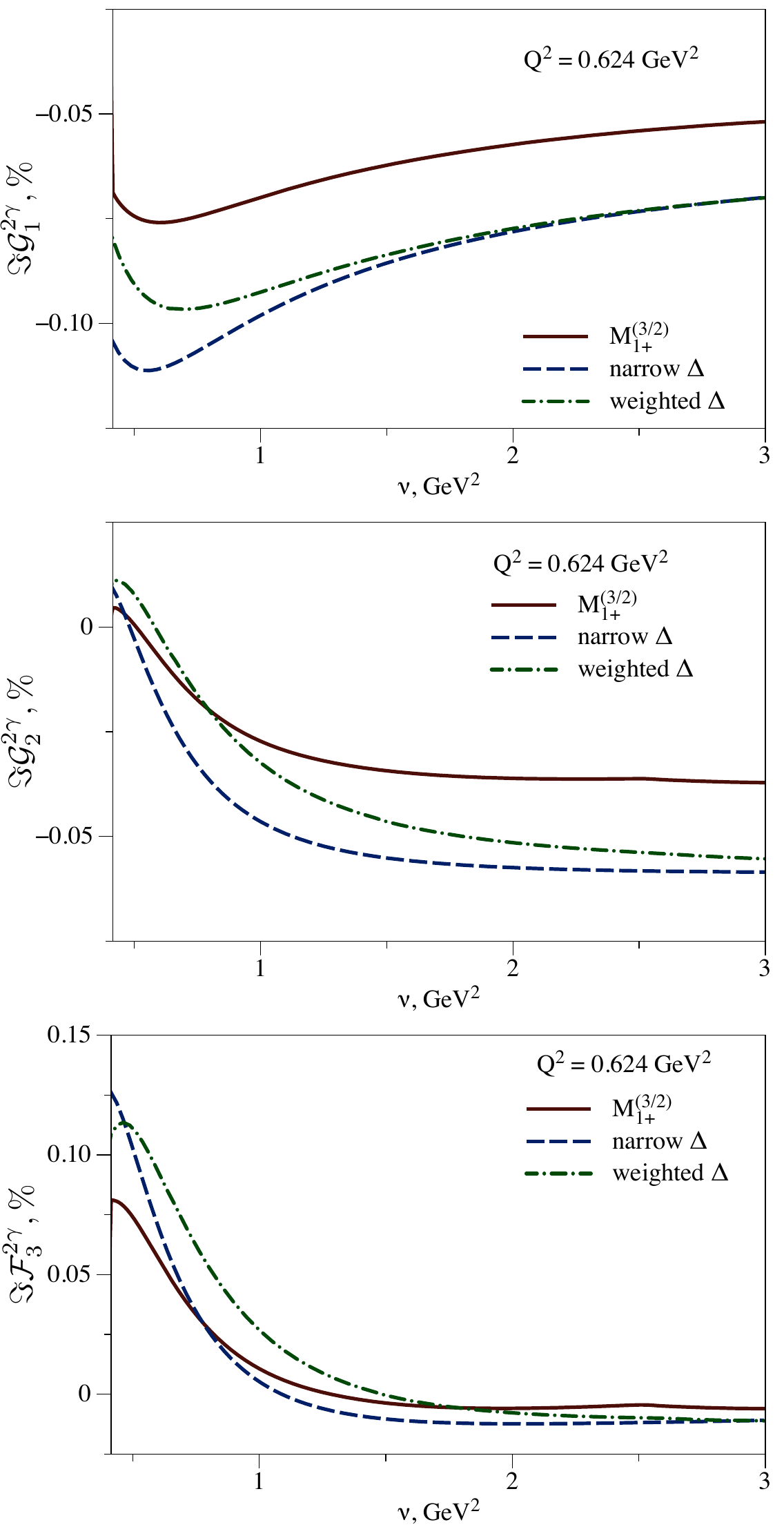}
\end{center}
\caption{Imaginary parts of the TPE amplitudes as a function of the crossing-symmetric variable $\nu$ for $ Q^2 = 0.624~\mathrm{GeV}^2$. The narrow- and weighted-$\Delta$ calculations are compared with the full $ M^{(3/2)}_{1+}$ $ \pi N$ multipole contribution, which is evaluated from the MAID 2007 fit.}
\label{imaginary_comparison_Q2_0_624_weight}
\end{figure}
We see from Fig. \ref{imaginary_comparison_Q2_0_624_weight} that at large values of $ \nu $, corresponding to higher energies, the weighted-$\Delta$ model gives a result similar to the narrow-$\Delta$ calculation. At lower $\nu$, corresponding to the $\Delta$-resonance region, the weighted-$\Delta$ calculation that accounts for the finite width effects is expected to be more realistic and shows differences from the narrow-$\Delta$ result. The leading $ M^{(3/2)}_{1+}$ $ \pi N$ contribution has a similar size, sign and behavior as the  model-$\Delta$ calculations. The difference is given mainly by the nonresonant background contributions which are included in the full $ M^{(3/2)}_{1+}$ multipole result.

\subsection{Dispersion relations at fixed $Q^2$ }
\label{sec33}

In this section, we perform the dispersion relation evaluation of the model-$\Delta$  TPE amplitudes and compare the results to the box graph model of Sec. \ref{sec31}.

The TPE amplitudes $ \cG^{2 \gamma}_M(\nu,Q^2), ~\cF^{2 \gamma}_2(\nu,Q^2), ~\cG^{2 \gamma}_1(\nu,Q^2), ~\cG^{2 \gamma}_2(\nu,Q^2) $ are odd functions under crossing $ \nu \to - \nu $, whereas the amplitude $ \cF^{2 \gamma}_3(\nu,Q^2) $ is even in $\nu$.  A general analysis of helicity amplitudes for the $ep \rightarrow ep$ process \cite{Kivel:2012vs} shows that in the Regge limit $\nu\rightarrow\infty$, $Q^{2}/\nu \rightarrow 0$ the functions ${\cal G}_{1,2},~{\cal F}_3$ vanish. Such high-energy behavior allows one to write down the following unsubtracted DRs at a fixed value of the momentum transfer $ Q^2 $ \cite{Borisyuk:2008es,Tomalak:2014sva,Tomalak:2016vbf}:
\ber
 \label{oddDR}
 \Re {\cal{G}}^{\mathrm{\mathrm{odd}}}(\nu, Q^2) & = & \frac{2 \nu}{\pi} \fint \limits^{~ \infty}_{\nu_{\mathrm{thr}}} \frac{\Im {\cal{G}}^{\mathrm{\mathrm{odd}}} (\nu^\prime, Q^2)}{{\nu^\prime}^2-\nu^2}  \mathrm{d} \nu^\prime, \\
 \label{evenDR}
 \Re  {\cal{F}}^{2 \gamma}_3 (\nu, Q^2) & = & \frac{2}{\pi} \fint \limits^{~ \infty}_{\nu_{\mathrm{thr}}}  \nu^\prime \frac{\Im  {\cal{F}}^{2 \gamma}_3  (\nu^\prime, Q^2)}{{\nu^\prime}^2-\nu^2}  \mathrm{d} \nu^\prime,
\eer
where $ {\cal{G}}^{\mathrm{\mathrm{odd}}}$ denotes any amplitude odd in $\nu$. The imaginary part in Eqs. (\ref{oddDR}), (\ref{evenDR}) is taken from the $s$-channel discontinuity only. These DRs are valid for the contribution of each intermediate state. In this section, we evaluate the dispersive integral for the narrow-$\Delta$ inelastic contribution, which starts from the $\Delta$-production threshold $ \nu_{\mathrm{thr}} = ( M^2_\Delta - M^2 )/ 2 - Q^2 / 4 $.

The unsubtracted DRs as given by Eqs. (\ref{oddDR}), (\ref{evenDR}) can only be written down for the functions with an appropriate HE behavior, when the contribution from the contour at infinity vanishes. We will next discuss the HE behavior of the TPE invariant amplitudes reconstructed within unsubtracted DRs and in the box graph model with the narrow-$\Delta$ intermediate state. 

First, we discuss the possible HE behavior of the amplitudes real parts reconstructed within the unsubtracted DRs of Eqs. (\ref{oddDR}), (\ref{evenDR}). We start with the case of the odd amplitude $ \cG^{\mathrm{odd}} $ and assume in the following the HE behavior of the imaginary part $ \Im \cG^{\mathrm{odd}} ( \nu,~Q^2) \simeq \nu^{\beta} \left( c_1 + c_2 \ln \nu + c_3 \ln^2 \nu \right) $ with the integer $ \beta \leq 0 $, which is sufficient for the convergence of the DR integral, keeping the squared logarithmic term as a Froissart bound \cite{Froissart:1961ux}. The corresponding exponent $ \tilde{\beta} $ in the HE behavior of the odd amplitude $ \Re {\cal{G}}^{\mathrm{odd}}(\nu, Q^2) \simeq \nu^{\tilde{\beta}} \left( \tilde{c}_1 + \tilde{c}_2 \ln \nu + \tilde{c}_3 \ln^2 \nu + \tilde{c}_4 \ln^3 \nu \right)$, which is reconstructed within the unsubtracted DR, in general has the upper bound $ \tilde{\beta} \leq -1 $ with the nonzero coefficients $ \tilde{c}_2,~\tilde{c}_3,~\tilde{c}_4$ only for $\beta = -1$ and can be constant (logarithmic) only in the case of the logarithmic leading behavior of the imaginary part $ \Im {\cal{G}}^{\mathrm{odd}}(\nu, Q^2) \sim  \ln \nu $ ($ \Im {\cal{G}}^{\mathrm{odd}}(\nu, Q^2) \sim  \ln^2 \nu $) respectively. We next turn to the even amplitude. Similarly, we assume the HE behavior of the imaginary part $ \Im \cF^{2 \gamma}_{3} ( \nu,~Q^2) \simeq \nu^{\beta} \left( c_1 + c_2 \ln \nu  + c_3 \ln^2 \nu \right) $ with the integer $ \beta \leq -1 $. The HE behavior of the real part of the even amplitude in the unsubtracted DR analysis is expected to be vanishing. In general, the corresponding exponent $ \tilde{\beta}$ in the HE behavior of the real part $ \Re \cF^{2 \gamma}_{3} (\nu, Q^2) \simeq \nu^{\tilde{\beta}} \left( \tilde{c}_1 + \tilde{c}_2 \ln \nu + \tilde{c}_3 \ln^2 \nu + \tilde{c}_4 \ln^3 \nu \right)$ has the upper bound $ \tilde{\beta} \leq -2 $ with nonzero $ \tilde{c}_2,~\tilde{c}_3,~\tilde{c}_4$ only for $\beta = -2$. The behavior of the imaginary part with $\beta = -1 $ is an exceptional case. The corresponding HE behavior of the real part $1/\nu$ ($\ln \nu / \nu$) is possible when the imaginary part behaves as $ \Im \cF^{2 \gamma}_{3}(\nu, Q^2) \sim  \ln \nu /\nu $ ($ \Im \cF^{2 \gamma}_{3} (\nu, Q^2) \sim  \ln^2 \nu /\nu^2 $) at high energies.

In the box graph model with the vertex of Eq. (\ref{real_delta}), the high-energy behavior of the TPE amplitudes is given by
\ber
\Im \cG^{2 \gamma}_M,~\Im \cF^{2 \gamma}_2,~\Im \cG^{2 \gamma}_1,~\Im \cG^{2 \gamma}_2 \sim \mathrm{const}, \qquad \Im \cF^{2 \gamma}_3 \sim \frac{1}{\nu}, \label{im_parts} \\
\Re \cF^{2 \gamma}_3 \sim \mathrm{const}, \qquad  \Re \cG^{2 \gamma}_M,~\Re \cG^{2 \gamma}_1,~\Re \cG^{2 \gamma}_2 \sim  \nu, \qquad  \Re \cF^{2 \gamma}_2 \sim \frac{\ln \nu}{\nu}. \label{re_parts} 
\eer
The behavior of Eq. (\ref{im_parts}) ensures the integrals in Eqs. (\ref{oddDR}), (\ref{evenDR}) are convergent for all amplitudes. 

However, the linear rise of the real parts $ \Re \cG^{2 \gamma}_1,~\Re \cG^{2 \gamma}_2 $ at high energies in the box graph model results in a linear growth of the TPE correction to the unpolarized cross section ($\delta_{2 \gamma}$) and to the polarization transfer ratio $ P_t/P_l$ of Eqs. (\ref{unpolarized_cross_section}), (\ref{polarization_observables1}):
\ber
\delta_{2\gamma} (\nu \to \infty) \sim \nu,\qquad \frac{P_t}{P_l}(\nu \to \infty) \sim \nu,
\eer
violating the unitarity conditions \cite{Zhou:2014xka,Blunden:2017nby,Tomalak_PhD}: 
\ber
 \delta_{2\gamma} (\nu \to \infty) \to 0, \qquad \frac{P_t}{P_l}(\nu \to \infty) \to  - \frac{2 M }{Q}  \frac{G_E}{G_M}.
\eer
In contrast, the HE behavior of the TPE amplitudes evaluated by unsubtracted DRs is in agreement with unitarity.

\newpage
\begin{figure}[H]
\begin{center}
\includegraphics[width=1.\textwidth]{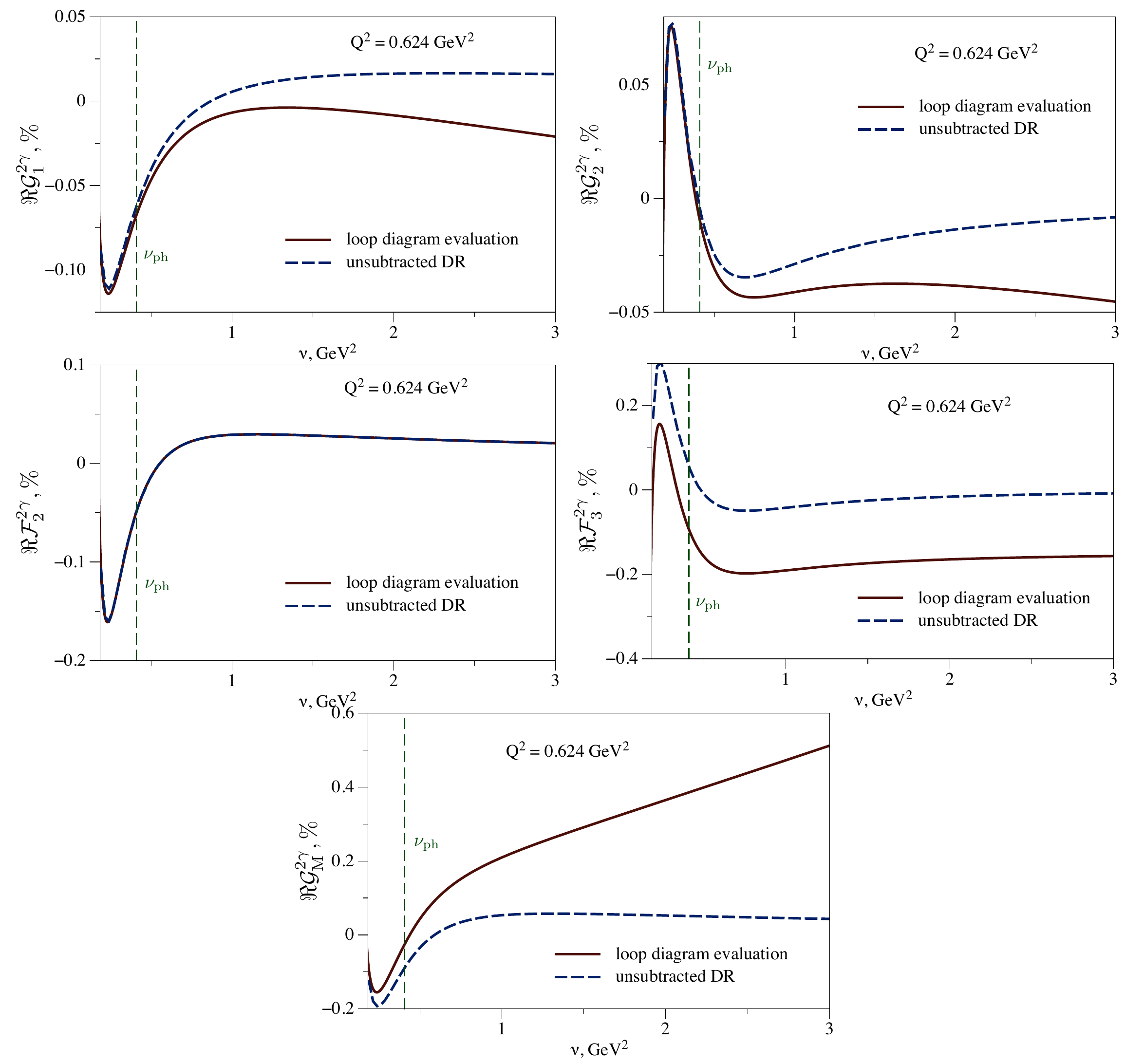}
\end{center}
\caption{Real parts of the TPE amplitudes as a function of the crossing-symmetric variable $\nu$ in the physical and unphysical regions for $ Q^2 = 0.624~\mathrm{GeV}^2$. We compare the direct loop diagram evaluation in the box graph model with narrow $\Delta$ to the calculation using unsubtracted dispersion relations. The vertical line corresponds with the boundary between physical and unphysical regions, i.e., $ \nu_{\mathrm{ph}} \approx 0.402 ~\mathrm{GeV}^2 $.}
\label{real_comparison_Q2_0_624}
\end{figure}
\begin{figure}[H]
\begin{center}
\includegraphics[width=1.\textwidth]{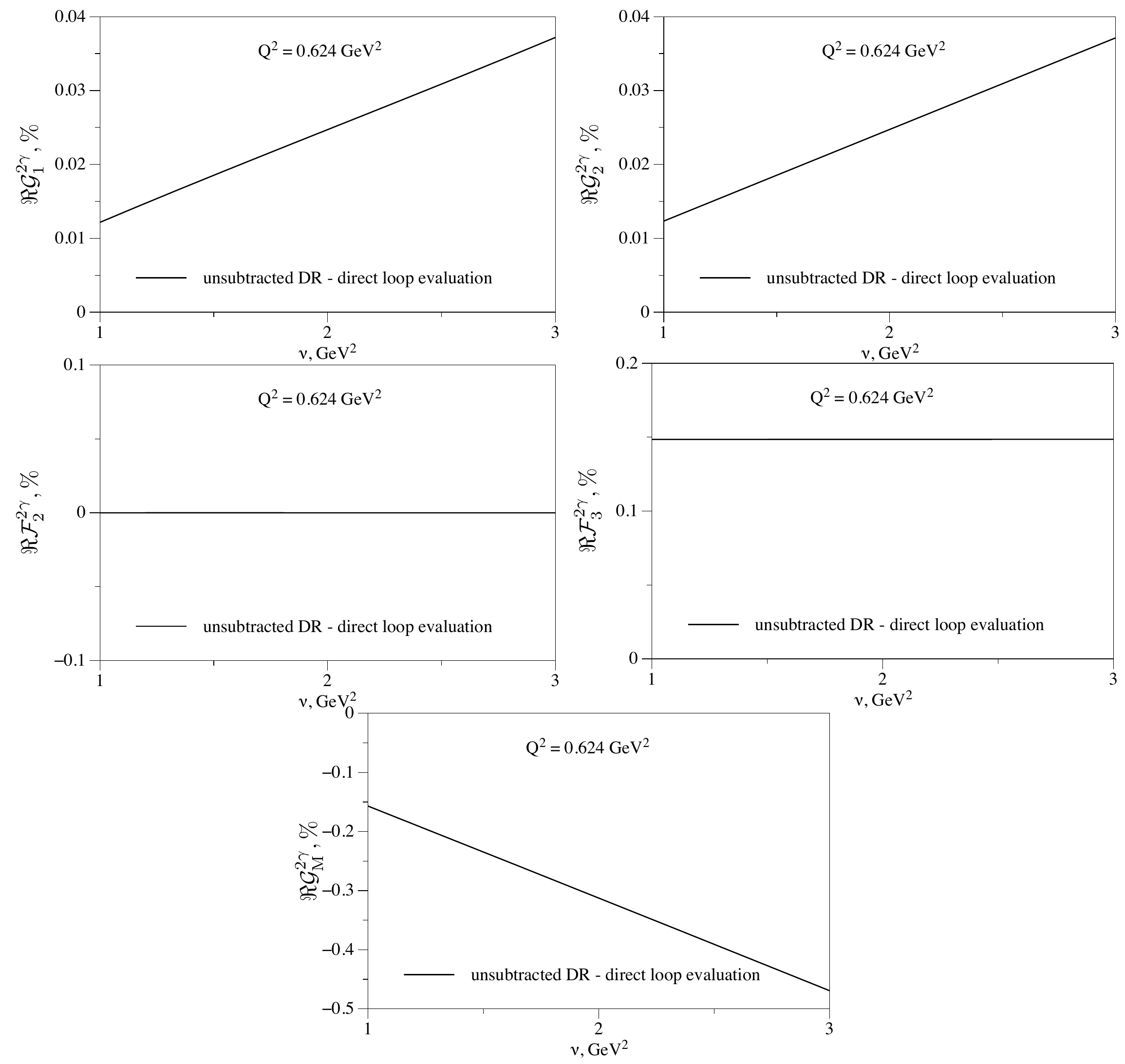}
\end{center}
\caption{The same as Fig. \ref{real_comparison_Q2_0_624}, but for the difference between the calculations from the unsubtracted DR and the direct loop diagram evaluation in the box graph model with narrow $\Delta$.}
\label{real_comparison_Q2_0_624_difference}
\end{figure}

In Fig. \ref{real_comparison_Q2_0_624}, we compare the box graph model result for the real part of the TPE amplitudes to the unsubtracted DRs result, see Eqs. (\ref{oddDR}), (\ref{evenDR}),  for $ Q^2 = 0.624~\mathrm{GeV}^2$. As the unsubtracted DR result is based on unitarity using on-shell input information only in evaluating the imaginary parts, and relies on analyticity to reconstruct the real parts, it gives the correct result for the TPE amplitudes. The direct loop diagram evaluation in the box graph model on the other hand, although based on the same on-shell input for the imaginary part is in general a model for the real part, as it makes an assumption on the vertices for off-shell kinematics. We notice from Fig. \ref{real_comparison_Q2_0_624} that only the amplitude $ \cF_2^{2\gamma}$ is correctly determined by the loop diagram evaluation in the box graph model. The results for other amplitudes are in clear disagreement. To provide more insights into these discrepancies, we show the difference between the unsubtracted DRs calculation and the loop diagram evaluation of the real parts within the box graph model in Fig. \ref{real_comparison_Q2_0_624_difference}. Figure \ref{real_comparison_Q2_0_624_difference} reveals that the difference between both ways of evaluating the odd amplitudes $\cG_1,~\cG_2$ and $\cG_M$ is a linear function in $ \nu$ and the difference for the even amplitude $ \cF_3^{2\gamma}$ is a constant. We also checked that when we perform one subtraction in the loop diagram evaluation all amplitudes in the box graph model agree with a once-subtracted DR, when choosing the same subtraction constant.

\subsection{Analytical continuation into the unphysical region}
\label{sec34}

In order to evaluate the dispersive integrals in Eqs. (\ref{oddDR}), (\ref{evenDR}) for the realistic $ \pi N $ intermediate state contribution, we need to know the imaginary parts of the invariant amplitudes from the threshold energy, corresponding to $ \nu_{\mathrm{thr}} = M m_\pi + m_\pi^2 / 2 - Q^2 / 4 $, upwards. The $ \pi N $ contribution was evaluated in Ref.~\cite{Tomalak:2016vbf} for the kinematics where only the input from the physical region of the $e p \to e p$ process is needed, which is possible when $ Q^2 < 0.064 ~ \mathrm{GeV}^2 $. At larger momentum transfers $ Q^2 > 0.064 ~ \mathrm{GeV}^2 $, the unphysical region starts to contribute to the dispersive integrals. In this section, we describe the procedure of analytical continuation of the imaginary parts of the $e p \to e p$ TPE amplitudes into the unphysical region from the knowledge of the amplitudes in the physical region. 

\begin{figure}[H]
\begin{center}
\includegraphics[width=0.75\textwidth]{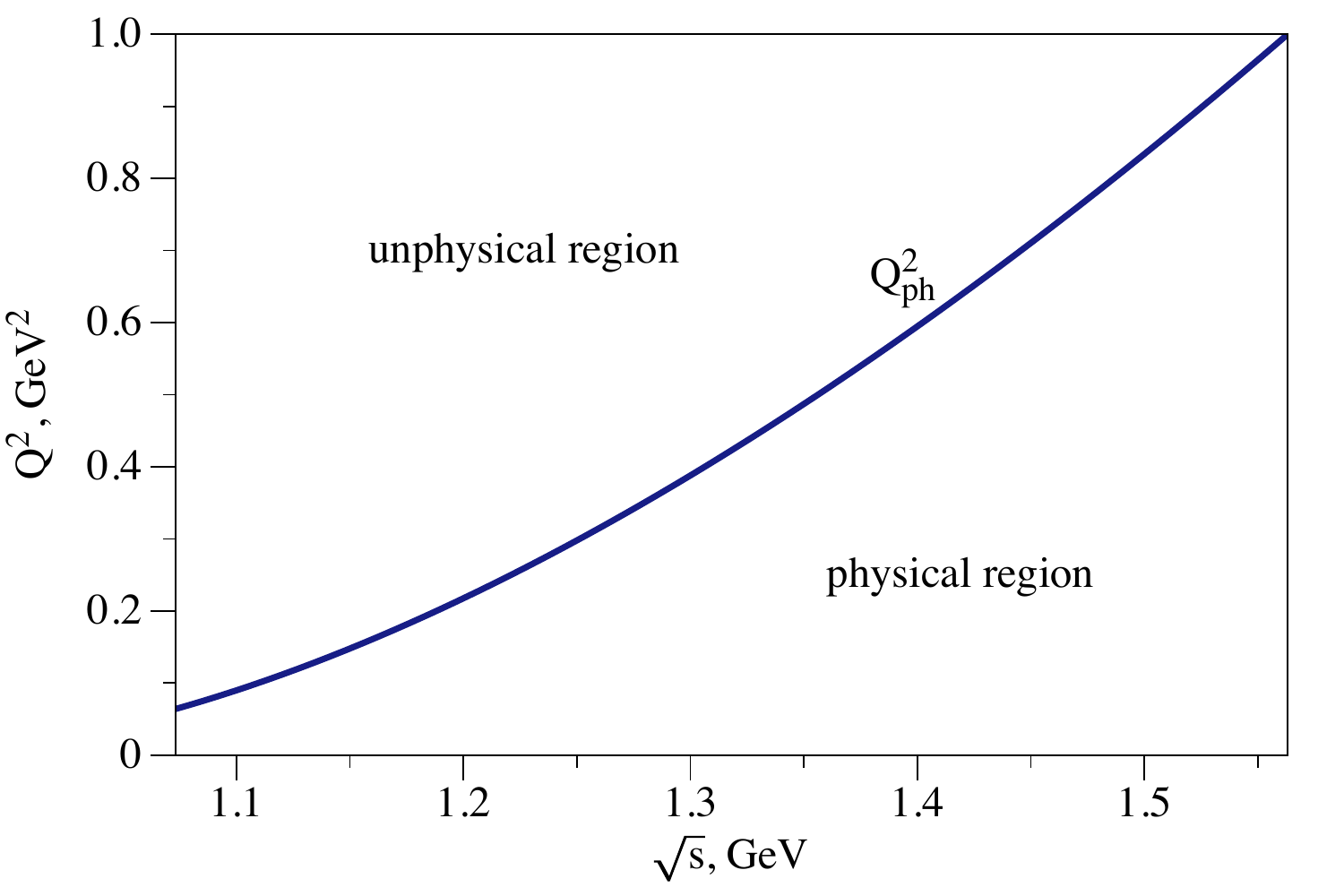}
\end{center}
\caption{Physical and unphysical regions of the TPE amplitudes in the ($Q^2,~\sqrt{s}$) plane of the $e p \to e p $ process. In the region $Q^2 > Q^2_{\mathrm{ph}} $, an analytical continuation into the unphysical region is required.}
\label{q2_cr_dependence_on_W}
\end{figure}

For a fixed value of $s$, the $e p \to e p$ TPE amplitudes will receive contributions which lie outside the physical region for $ Q^2 \ge Q^2_{\mathrm{ph}} \equiv \left( s - M^2 \right)^2/ s$. The boundary curve between physical and unphysical regions for the TPE amplitudes is shown in Fig. \ref{q2_cr_dependence_on_W} in the $(Q^2,~\sqrt{s})$ plane. We notice that at relatively small momentum transfer values $ 0.064 ~ \mathrm{GeV}^2< Q^2 \lesssim 0.5$-$0.6 ~ \mathrm{GeV}^2$, the dominant contribution from the unphysical region entering the dispersive integrals originates from the $\pi N$ threshold and the $ \Delta$-resonance regions. Consequently, the procedure of analytical continuation can be developed and tested on the example of the model-$\Delta $ calculation of Secs. \ref{sec31}-\ref{sec33}, where we know the imaginary parts exactly in both physical and unphysical regions from the direct loop diagram evaluation. In order to correctly reproduce the position of the inelastic $ \pi N $ cut and to qualitatively account for the $ \Delta$-resonance width, we study the weighted-$\Delta$ TPE correction with the weighting function of Eq. (\ref{narrow_delta_weight}).

First, we evaluate the imaginary parts of the $ e p \to e p $ scattering amplitudes in the physical region for a fixed value of $s$, corresponding to a fixed value of the lepton beam energy in the lab frame, as a function of $Q^2$ by using the unitarity relations \cite{Pasquini:2004pv,Tomalak:2016vbf}. We then fit, for a fixed value of $s$, the obtained $Q^2$ dependence by a sum of the leading terms in the $Q^2$ expansion of the inelastic TPE amplitudes \cite{Brown:1970te,Gorchtein:2014hla,Tomalak:2015aoa,Tomalak_PhD}:
\ber
\Im \cG^{2 \gamma}_1 \left(s,~Q^2 \right) & \sim & Q^2 f \left(s,~Q^2 \right), \label{amplitude_g1} \\
\Im \cG^{2 \gamma}_2 \left(s,~Q^2 \right)  & \sim & Q^2 f \left(s,~Q^2 \right), \\
\Im \cF^{2 \gamma}_3 \left(s,~Q^2 \right)  & \sim & f \left(s,~Q^2 \right),  \label{amplitude_f3}
\eer
with a form for the fitting function:
\ber \label{fitting_function}
f (s,~Q^2) \equiv a_1( s ) + a_2( s )  \ln Q^2 + a_3( s )  Q^2 + a_4( s )  Q^2 \ln Q^2 + a_5( s )  Q^4 + a_6( s )  Q^4 \ln Q^2.
\eer 
The fit coefficients $a_1(s)$, ..., $a_6(s)$ at a fixed value of $s$ are obtained for each amplitude separately. 
For relatively small values of the c.m. energy, slightly above the pion-production threshold $ s_{\mathrm{thr}} = 1.152~\mathrm{GeV}^2 $, i.e. $ s_{\mathrm{thr}} \leq s \lesssim 1.3~\mathrm{GeV}^2$, the fit for all amplitudes $\Im \cG^{2 \gamma}_1$, $\Im \cG^{2 \gamma}_2$ and $\Im \cF^{2 \gamma}_3$ is well described by two coefficients $a_1( s ),~a_2( s )$ only. For larger values of $s$: $ 1.3~\mathrm{GeV}^2 \lesssim s $ for $\Im \cG^{2 \gamma}_1$, $  1.3~\mathrm{GeV}^2 \lesssim s \lesssim 1.9 ~\mathrm{GeV}^2$ for $\Im \cG^{2 \gamma}_2$, and $  1.3~\mathrm{GeV}^2 \lesssim s \lesssim 1.8 ~\mathrm{GeV}^2$ for $\Im \cF^{2 \gamma}_3$ we perform a four-parameter fit with coefficients $a_1( s ),~a_2( s ),~a_3( s ),~a_4( s )$. For even larger c.m. energies, i.e. $ s \gtrsim 1.9 ~\mathrm{GeV}^2$ for $\Im \cG^{2 \gamma}_2$, and $ s \gtrsim 1.8 ~\mathrm{GeV}^2$ for $\Im \cF^{2 \gamma}_3$, we use the six-parameter functional form of Eq. (\ref{fitting_function}).

\begin{figure}[ht]
\begin{center}
\includegraphics[width=1.\textwidth]{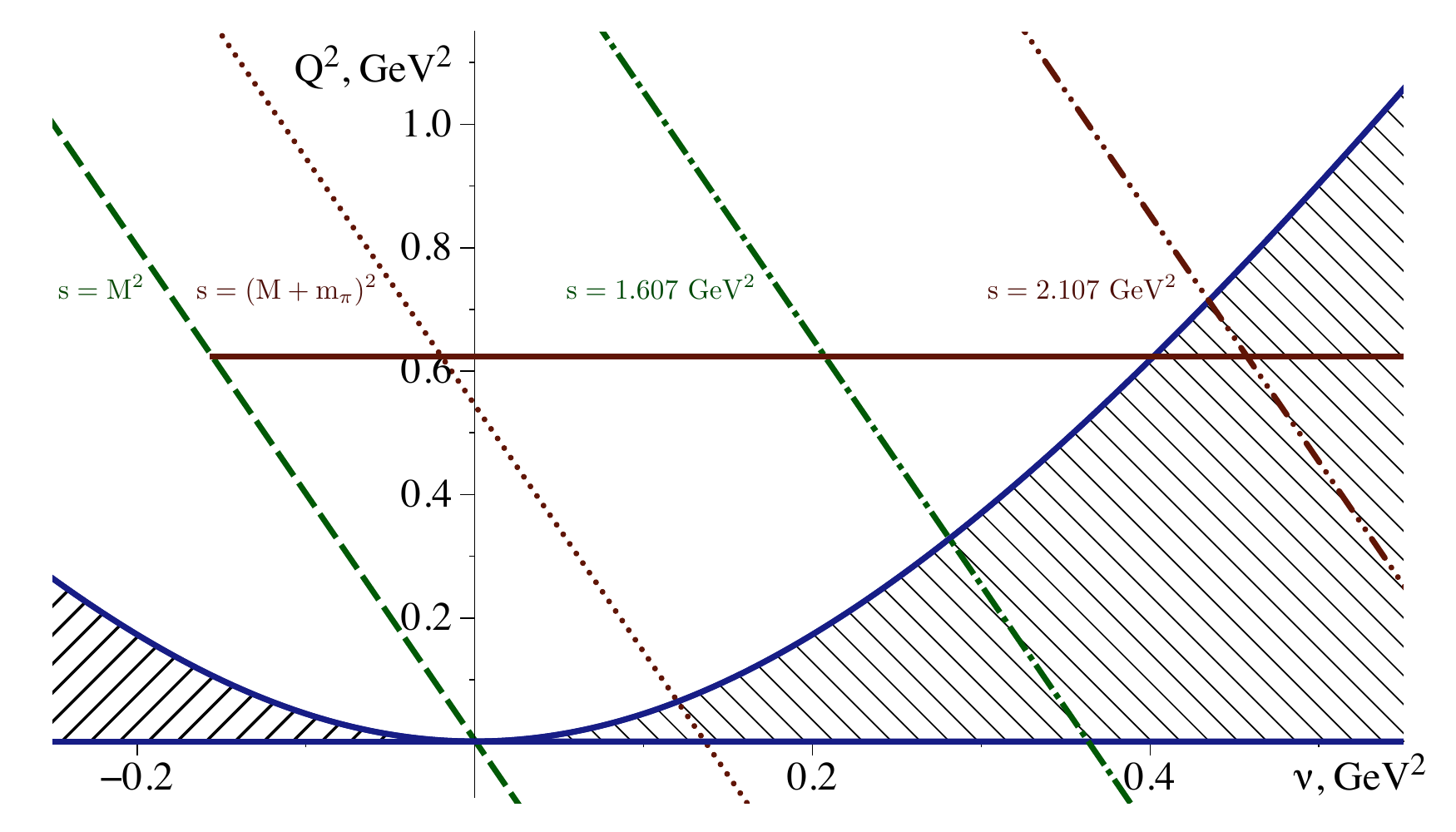}
\end{center}
\caption{Physical and unphysical regions of the kinematical variables $ \nu $ and $ Q^2 $ (Mandelstam plot) for the elastic electron-proton scattering. The hatched blue region corresponds to the physical region, the green-dashed and red-dotted lines give the elastic and the pion-nucleon ($\pi N$) threshold positions in the $s$ channel, the green dashed-dotted and red dashed-double-dotted lines correspond with the threshold positions in the $s$ channel of the states with the invariant masses $ W^2 = 1.607~\mathrm{GeV^2}$ and $ W^2 =  2.107~\mathrm{GeV^2} $ respectively. The horizontal red curve at fixed $Q^2=0.624~\mathrm{GeV}^2$ illustrates the path of the dispersive integral corresponding with the kinematics of Fig. \ref{real_comparison_Q2_0_624}.}
\label{regions_electron}
\end{figure}

In the following we test this procedure of analytical continuation for two values of $s$ as a function of $Q^2$. The physical and unphysical regions for these two $s$ values are visualized in Fig. \ref{regions_electron}. 
\begin{figure}[h!]
\begin{center}
\includegraphics[width=0.9\textwidth]{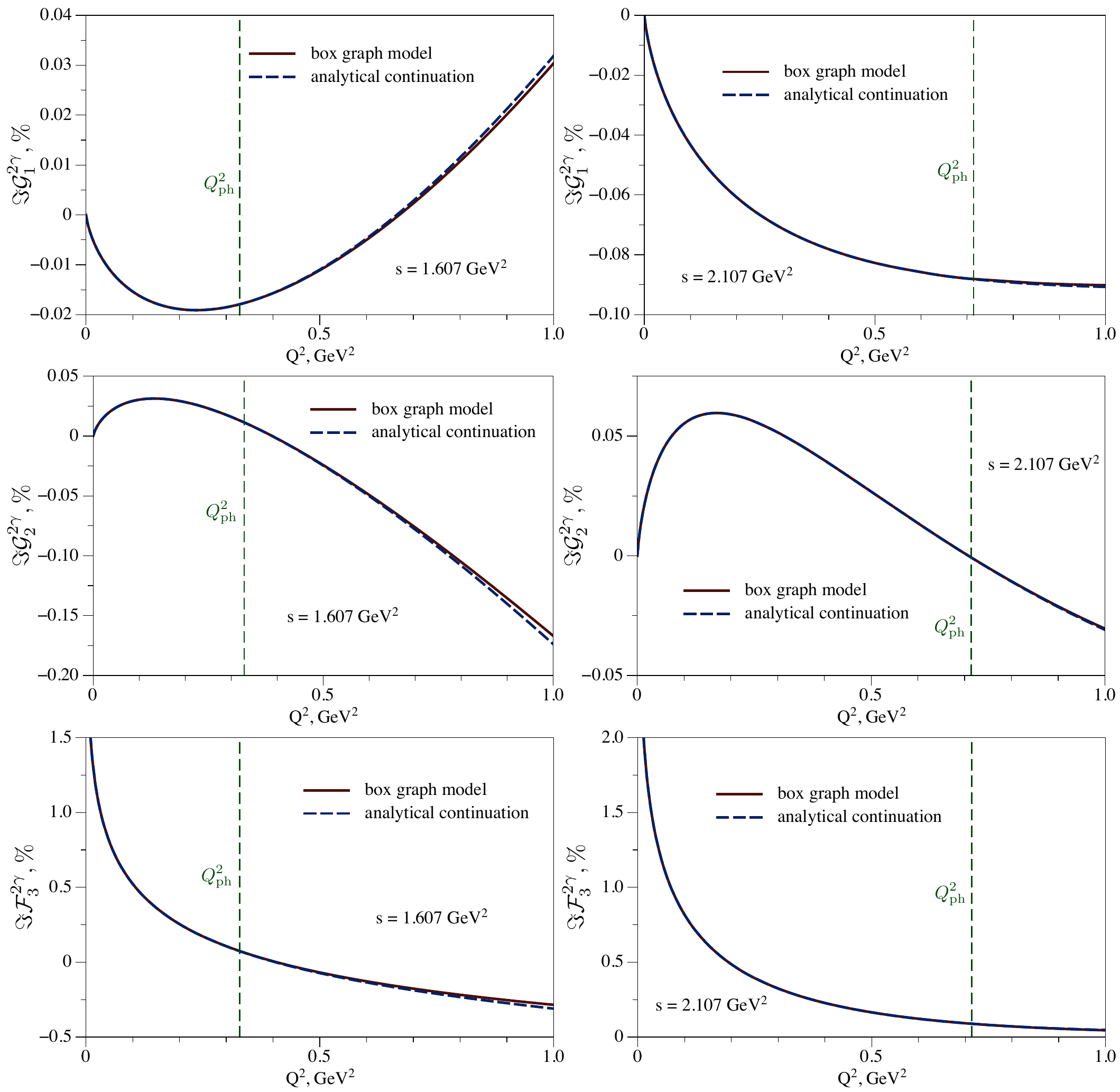}
\end{center}
\caption{The imaginary parts of the TPE amplitudes $ \cG^{2 \gamma}_1,~\cG^{2 \gamma}_2,~\cF^{2 \gamma}_3 $ which are reconstructed by the fits of Eqs. (\ref{amplitude_g1})-(\ref{amplitude_f3}) in comparison with the results from the direct loop diagram evaluation in the box graph model with weighted $\Delta$ for the c.m. squared energies $ s = 1.607~\mathrm{GeV}^2$ (left panel) and $ s = 2.107~\mathrm{GeV}^2$ (right panel). The vertical lines correspond with the boundary between physical and unphysical regions, i.e., $ Q^2_{\mathrm{ph}} \approx 0.329 ~\mathrm{GeV}^2 $ (left panel), and $ Q^2_{\mathrm{ph}} \approx 0.714 ~\mathrm{GeV}^2 $ (right panel).}
\label{imaginary_reconstruction}
\end{figure}

In Fig. \ref{imaginary_reconstruction}, we compare, for two values of $s$, the analytical continuation of the imaginary part of the TPE amplitudes as reconstructed from the amplitudes in the physical region only by the fits of Eqs. (\ref{amplitude_g1})-(\ref{amplitude_f3}) with the box graph model with a $\Delta$-intermediate state weighted by the function of Eq. (\ref{narrow_delta_weight}). We notice a very good agreement between both calculations up to $Q^2$ values of at least $Q^2 = 1~\mathrm{GeV}^2$.

\begin{figure}[ht]
\begin{center}
\includegraphics[width=0.99\textwidth]{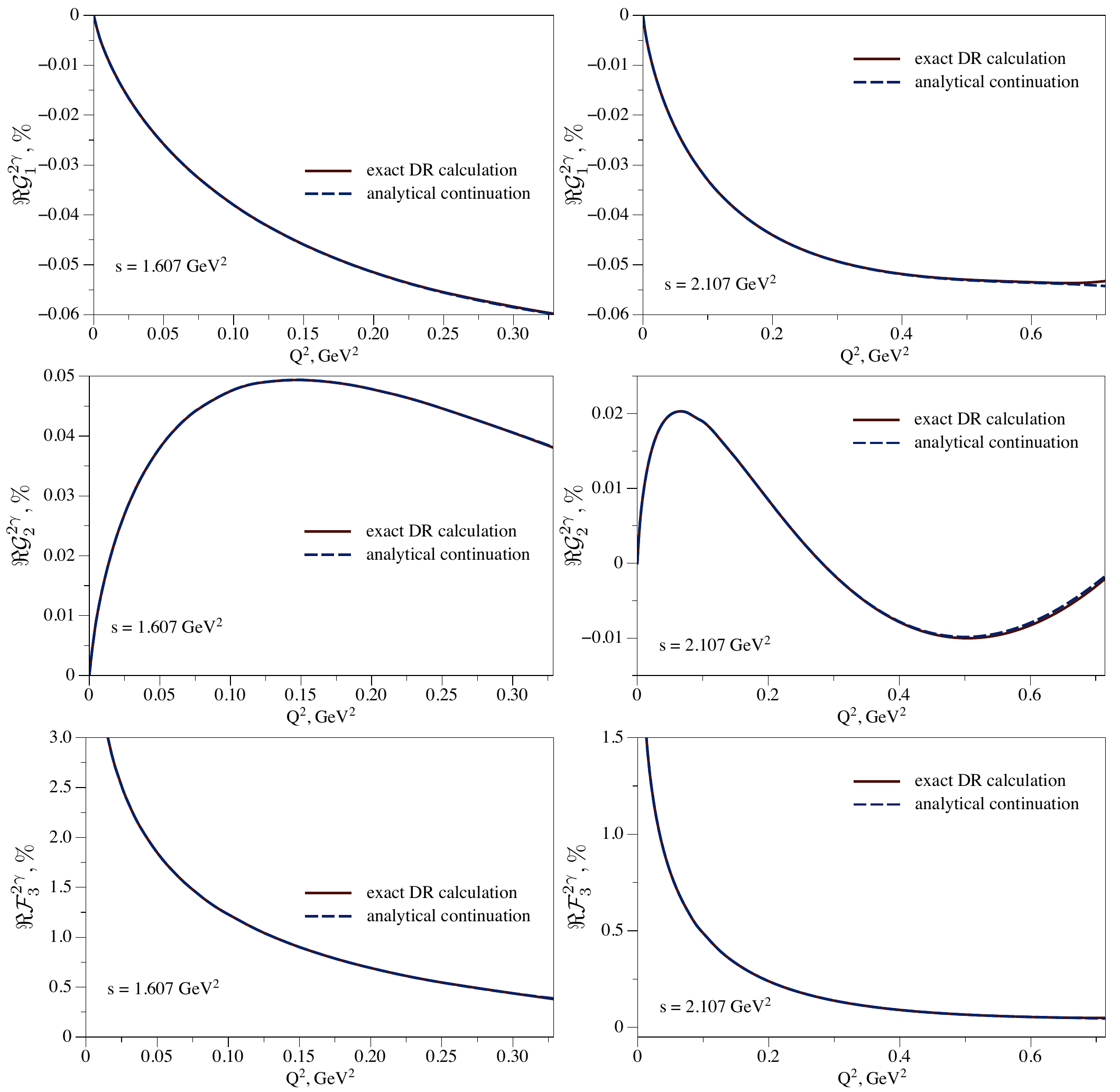}
\end{center}
\caption{Same as Fig. \ref{imaginary_reconstruction}, but for the real parts of the TPE amplitudes in the physical region.}
\label{real_reconstruction}
\end{figure}

Using the imaginary parts of the TPE amplitudes evaluated either from the exact direct loop diagram calculation in the box graph model, or from the analytical continuation described above, we next perform the the dispersion integrals of Eqs. (\ref{oddDR}), (\ref{evenDR}) to obtain the real parts of the TPE amplitudes. In Fig. \ref{real_reconstruction}, we present the thus obtained real parts of the TPE amplitudes in the physical region. We notice from Fig. \ref{real_reconstruction} that both ways of evaluating the real parts are in a very good agreement over the whole physical region of the scattering process. 
\begin{figure}[H]
\begin{center}
\includegraphics[width=0.5\textwidth]{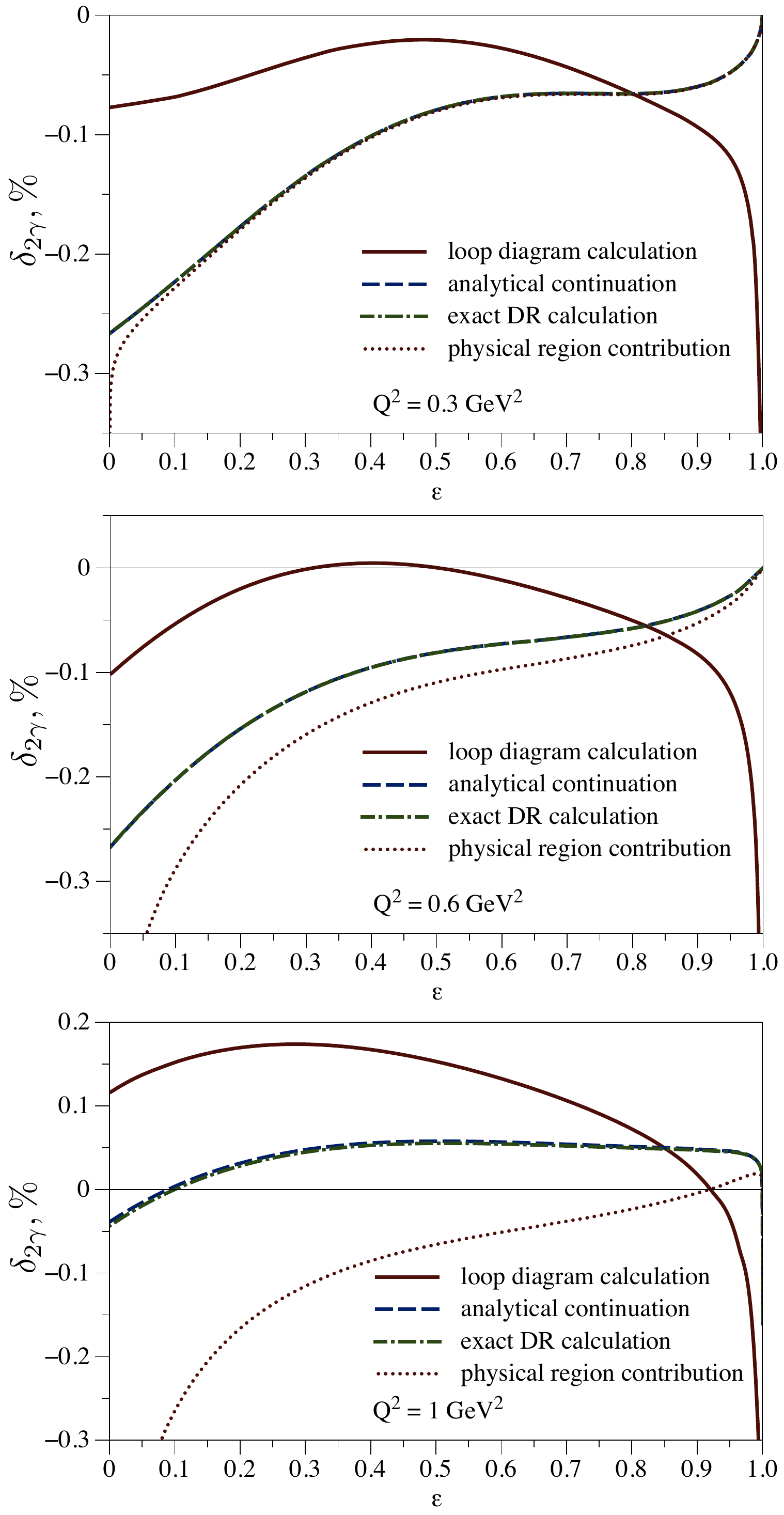}
\end{center}
\caption{The inelastic TPE correction of Eq. (\ref{unpolarized_cross_section}) to the unpolarized $ e^{-} p \to e^{-} p$ cross section for three values of $Q^2$ evaluated within unsubtracted DRs from the analytically continued imaginary parts of the TPE amplitudes in comparison with the similar evaluation (exact DR calculation), when the box graph model with weighted $\Delta$ is used to obtain the imaginary parts of the TPE amplitudes in the unphysical region. We furthermore compare with the contribution from the physical region only, as well as with the direct loop diagram evaluation of the real parts in the box graph model with weighted $\Delta$, as outlined in Sec. \ref{sec31}. Upper plot: $Q^2 = 0.3~\mathrm{GeV}^2$, central plot: $Q^2 = 0.6~\mathrm{GeV}^2$, lower plot: $Q^2 = 1~\mathrm{GeV}^2$.}
\label{cross_section_reconstruction}
\end{figure}

Substituting the real parts into the cross section correction expression of Eq. (\ref{unpolarized_cross_section}) we show our results for this observable in Fig. \ref{cross_section_reconstruction}. The comparison between the analytically continued evaluation with the exact realization of the dispersion relations (exact DR calculation), when taking the imaginary parts of the box graph model with weighted $\Delta$ both in physical and unphysical regions as an input, indicates the very good quality of the method described above. To illustrate the growing importance of the unphysical region contribution with increasing momentum transfer $Q^2$, we also show the TPE correction coming from the physical region only in Fig. \ref{cross_section_reconstruction}. The unphysical region contribution becomes more important for backward scattering kinematics, i.e. smaller $\varepsilon$ values. In the vicinity of $\varepsilon = 0$, the cancellation of two infinitely large contributions from the physical and unphysical regions takes place. Furthermore, we display the results from the direct loop diagram calculation of the real parts of the TPE amplitudes in the box graph model taking the $\Delta$-intermediate state weighted by the function in Eq. (\ref{narrow_delta_weight}). As it was discussed in Sec. \ref{sec33}, the results of the loop diagram evaluation for the real parts differ from the dispersive evaluation and violate unitarity at high energies. For $ \varepsilon \to 1 $ the cross section correction diverges as $ \delta_{2 \gamma} \to 1/\sqrt{1 - \varepsilon}$ in the box graph model.

\newpage
\section{$\pi N $ contribution}
\label{sec4}

In this section, we generalize the method of analytical continuation of Sec. \ref{sec34} to the case of the $ \pi N$ intermediate state contribution relying directly on the empirical information from the MAID 2007 fit~\cite{Drechsel:1998hk,Drechsel:2007if}.

\begin{figure}[H]
\begin{center}
\includegraphics[width=0.96\textwidth]{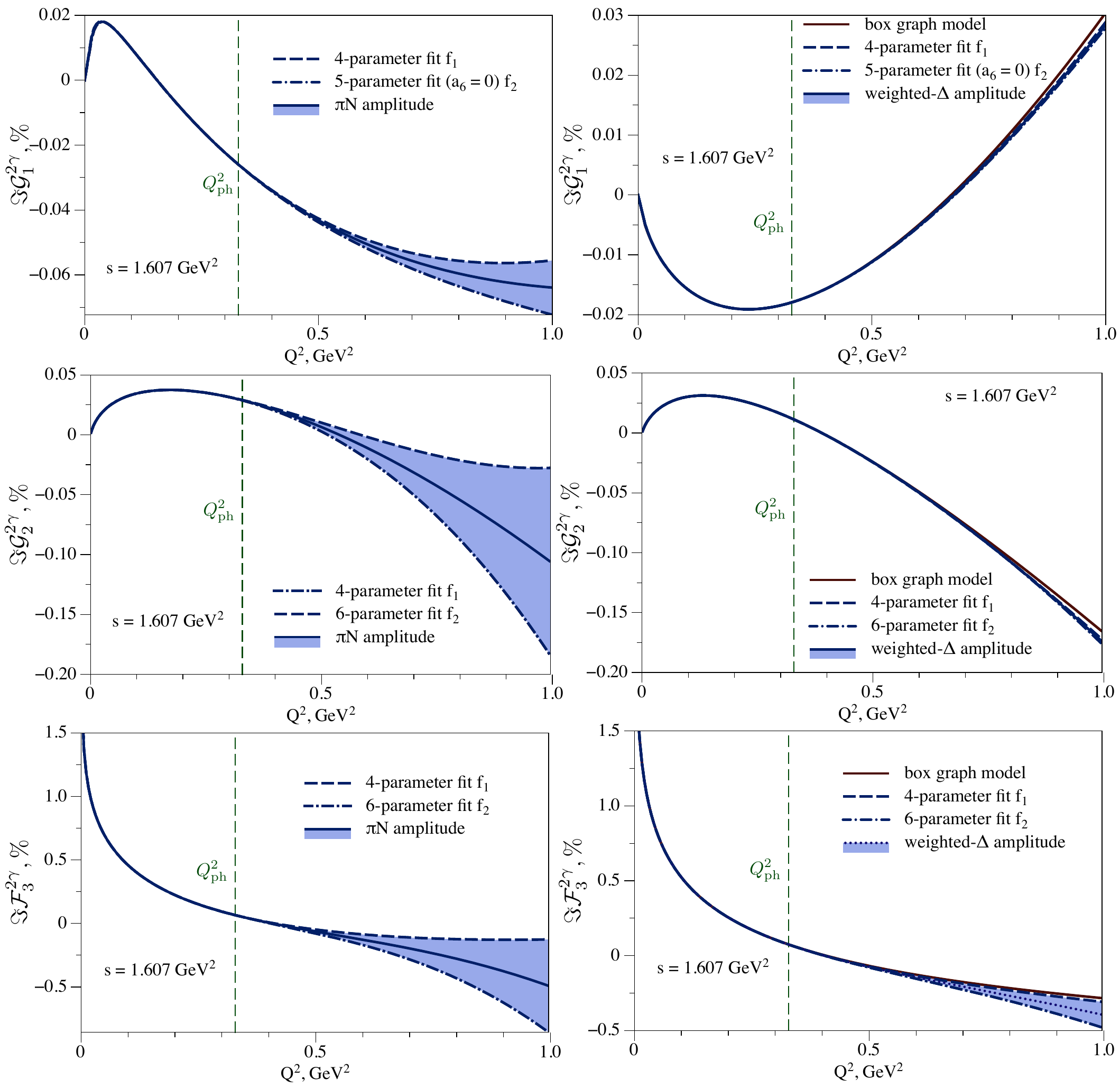}
\end{center}
\caption{The imaginary parts of the TPE amplitudes $ \cG^{2 \gamma}_1,~\cG^{2 \gamma}_2,~\cF^{2 \gamma}_3 $ from the $ \pi N $ (left panel) and weighted-$\Delta$ (right panel)  intermediate state contributions as reconstructed from fits of Eq. (\ref{weights}) for the c.m. squared energy $ s = 1.607~\mathrm{GeV}^2$. The analytical continuation of the $\Delta$-intermediate state amplitudes is compared with the exact result in the box graph model.  The vertical lines correspond with the boundary between the physical ($Q^2 < Q^2_{\mathrm{ph}}$) and unphysical ($Q^2 > Q^2_{\mathrm{ph}}$) regions: $ Q^2_{\mathrm{ph}} \approx 0.329 ~\mathrm{GeV}^2 $.}
\label{imaginary_reconstruction_fits}
\end{figure}

As a guiding principle, we exploit the general form of Eq. (\ref{fitting_function}) to describe the $Q^2$ dependence of the TPE amplitudes at a fixed value of $s$ and require the imaginary part of the invariant amplitudes to vanish at threshold, i.e. $ \Im \cG^{2\gamma} (s = s_\mathrm{thr},~Q^2) = 0$, where $ \cG$ stands for $\cG_1,~\cG_2$ or $\cF_3$. In Fig. \ref{imaginary_reconstruction_fits}, we compare the $ \pi N$ intermediate state contribution to the imaginary parts of the TPE amplitudes with the weighted-$\Delta$ model of Sec. \ref{sec3}. For the $\pi N$ intermediate state, we expect to have a similar shape as in the weighted-$ \Delta$ model calculation for the amplitudes $ \Im \cG^{2\gamma}_2 $ and $ \Im \cF^{2\gamma}_3$, as can be seen from the physical region in Fig. \ref{imaginary_reconstruction_fits}.

For the $\pi N$ intermediate state contribution we have no exact calculation to compare with when extrapolating into the unphysical region. We will therefore estimate the theoretical error of such extrapolation procedure by performing two different fits, labeled by $f_1$ and $f_2$. The TPE amplitudes $ \Im \cG^{2\gamma}$ are then given by
\ber \label{weights}
\Im \cG^{2\gamma} \left( s,~Q^2 \right)= \frac{ f_1 \left(s,~Q^2 \right) + f_2 \left(s,~Q^2 \right)}{2} \pm \frac{ | f_1 \left(s,~Q^2 \right) - f_2 \left(s,~Q^2 \right)|}{2} ,
\eer
where $f_1$ and $f_2$ have functional forms as in Eq. (\ref{fitting_function}) with a different number of nonzero parameters. The difference between both fits $f_1$ and $f_2$ in Eq. (\ref{weights}) will define our theoretical error band. We illustrate this procedure in Fig. \ref{imaginary_reconstruction_fits} for c.m. squared energy $s=1.607~\mathrm{GeV}^2$. For comparison, we also provide the same realization for the case of the weighted-$\Delta$ intermediate state.
In the following, we detail the form of the fit functions for the TPE amplitudes $\Im \cG^{2\gamma}_1,~\Im \cG^{2\gamma}_2,~\Im \cF^{2\gamma}_3$.

For the invariant amplitude $ \Im \cG^{2\gamma}_2$ in the region between the threshold and $\Delta$ peak position, when $s_{\mathrm{thr}} \leq s \lesssim 1.5~\mathrm{GeV}^2$, we use a two-parameter functional form (with $a_3 = a_4 = a_5 = a_6 = 0$), and a four-parameter functional form (with $a_5 = a_6 = 0$) in Eq. (\ref{weights}). For $ s \gtrsim 1.5~\mathrm{GeV}^2$, we use four-parameter (with $a_5 = a_6 = 0$) and six-parameter fits in order to have a similar $Q^2$ dependence as in the weighted-$\Delta$ model calculation.

We next describe the fits for the imaginary part of the amplitude $\Im \cF_3$. In order to satisfy simultaneously the vanishing behavior near the threshold, when $ s_{\mathrm{thr}} \leq s \lesssim 1.38~\mathrm{GeV}^2$, and to have a good description of the physical region, we use for $\Im \cF_3$ a three-parameter fit (with $a_3=a_5=a_6=0$), and another three-parameter fit (with $a_4=a_5=a_6=0$). For larger values of the c.m. energy ($ s > 1.38~\mathrm{GeV}^2$), we describe the amplitude $ \Im \cF_3^{2 \gamma}$ by six-parameter and four-parameter (with $a_5 = a_6 = 0$) fits.

We notice from Fig. \ref{imaginary_reconstruction_fits}, that the imaginary part of the amplitude $ \Im \cG^{2\gamma}_1$ for the $\pi N$ intermediate state has a different $Q^2$ dependence in the physical region as compared to the weighted-$\Delta$ model. The difference originates from the contributions of Born and vector meson terms in the $\pi N$ multipoles from the MAID fit as well as the subsequent unitarization of these multipoles.  Consequently, we cannot fully rely on the weighted-$\Delta$ model reconstructing the amplitude $ \Im \cG^{2\gamma}_1$. For small values of the c.m. energy $  s_{\mathrm{thr}} \leq s \lesssim 1.5~\mathrm{GeV}^2$, we choose a four-parameter form (with $a_5 = a_6 = 0$) of the fitting function without $Q^2$ multiplier in Eq. (\ref{amplitude_g1}), and for $ 1.5~\mathrm{GeV}^2 \lesssim s \lesssim 1.75~\mathrm{GeV}^2$ a five-parameter form ($a_6 = 0$) without $Q^2$ multiplier in Eq. (\ref{amplitude_g1}), and start fitting from $ Q^2 = 0.06~\mathrm{GeV}^2$ ($Q^2 = 0.03~\mathrm{GeV}^2$) respectively. In the region $ 1.75~\mathrm{GeV}^2 \lesssim s \lesssim 2~\mathrm{GeV}^2$ ($ s \gtrsim 2~\mathrm{GeV}^2 $), we choose six-parameter and four-parameter forms of the fitting function without $Q^2$ multiplier in Eq. (\ref{amplitude_g1}) and use the numerical evaluations as fit input starting from $ Q^2 = 0.06~\mathrm{GeV}^2$ ($Q^2 = 0$) respectively.

The error bands resulting from the difference between the two fits for $ \Im \cG_1,~\Im \cG_2$ and $\Im \cF_3$ are shown in Fig. \ref{imaginary_reconstruction_fits} for a value $s=1.607~\mathrm{GeV}^2$, slightly above the $\Delta$-resonance position.

\begin{figure}[H]
\begin{center}
\includegraphics[width=1.\textwidth]{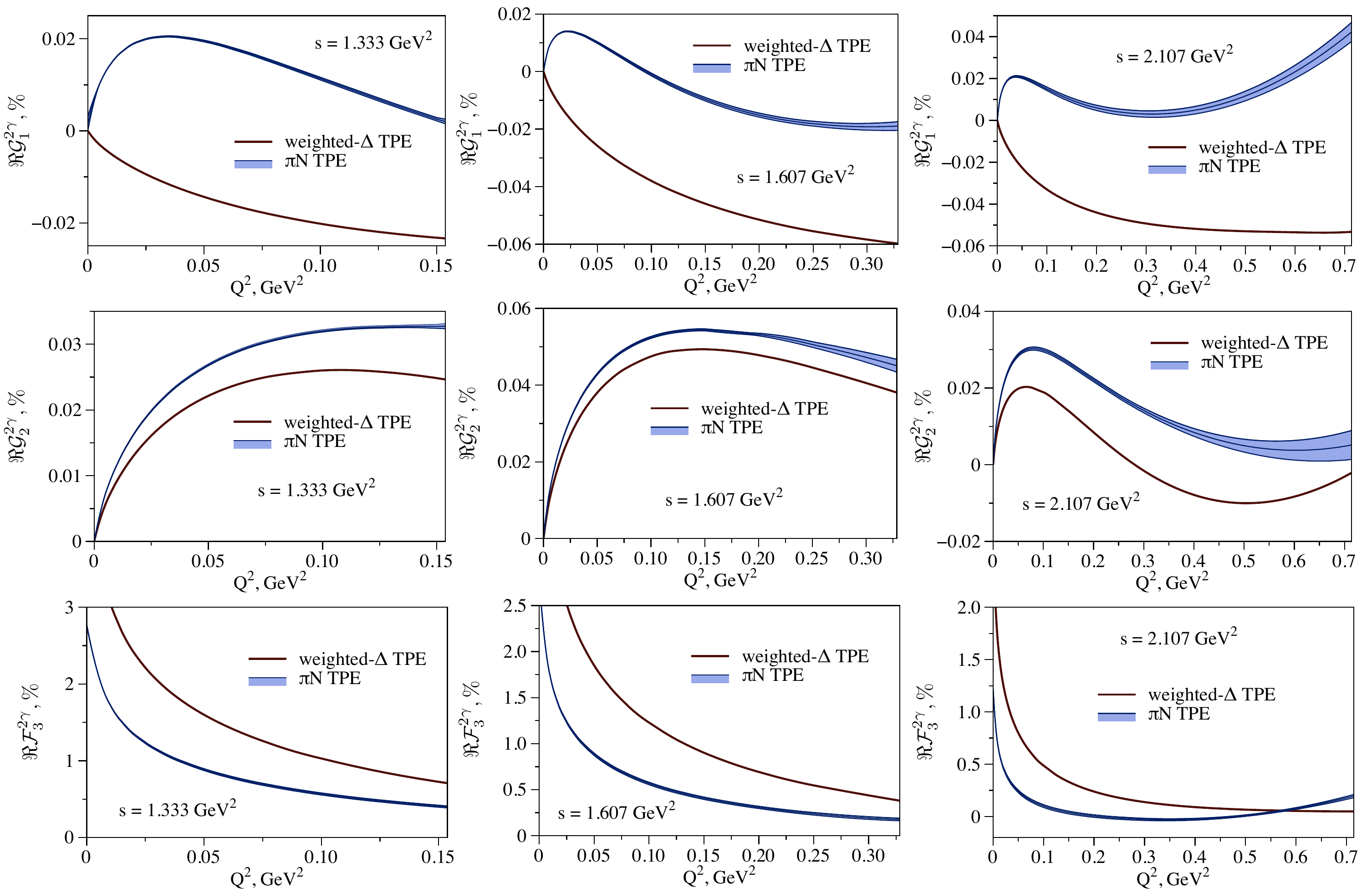}
\end{center}
\caption{The real parts of the TPE amplitudes $ \cG^{2 \gamma}_1,~\cG^{2 \gamma}_2,~\cF^{2 \gamma}_3 $ from the $ \pi N $ intermediate state contribution as reconstructed from fits of Eqs. (\ref{amplitude_g1})-(\ref{amplitude_f3}), in comparison with the weighted-$\Delta$ box graph model result for the c.m. squared energies $ s = 1.333~\mathrm{GeV}^2$ (left panel), $ s = 1.607~\mathrm{GeV}^2$ (middle panel) and $ s = 2.107~\mathrm{GeV}^2$ (right panel). The kinematical coverage corresponds to the physical region.}
\label{real_reconstruction_piN}
\end{figure}

Besides the uncertainty from the fit forms used in the analytical continuation, the second largest uncertainty comes from the region of large $W$. The MAID2007 fit~\cite{Drechsel:1998hk,Drechsel:2007if} is available for $ W  < 2.5~\mathrm{GeV}$ and qualitatively describes resonances and background up to $ W_0 = 2~\mathrm{GeV}$. We exploit the MAID parametrization up to $W_0$ and subsequently connect the end point of the $W$-integrand $F(W)$ to two functional forms:

\ber 
F \left( W \right) \mathrm{d} W & = & \frac{W_0^2}{W^2} F \left( W_0 \right)  \mathrm{d} W, \label{f_1}  \\
F \left( W \right) \mathrm{d} W & = & \frac{1 + e^{\frac{W_0 - W_1}{a}}}{1 + e^{\frac{2 W - W_0 - W_1}{a}}} F \left( W_0 \right)  \mathrm{d} W,  \label{f_2}
\eer
with $ W_1 = 3~\mathrm{GeV}$ and $ a = 0.5~\mathrm{GeV}$. We take the calculation with the integrand of Eq. (\ref{f_1}) as a central value, and estimate the uncertainty coming from the large-$W$ region as the difference between the results of Eqs. (\ref{f_1}) and (\ref{f_2}). We add the errors from the analytical continuation procedure and resulting from the large-$W$ extrapolation in quadrature. In Fig. \ref{real_reconstruction_piN}, we present the real part of the TPE amplitudes in the physical region and compare them to the dispersive evaluation of the weighted-$\Delta$ model. 

\begin{figure}[H]
\begin{center}
\includegraphics[width=0.47\textwidth]{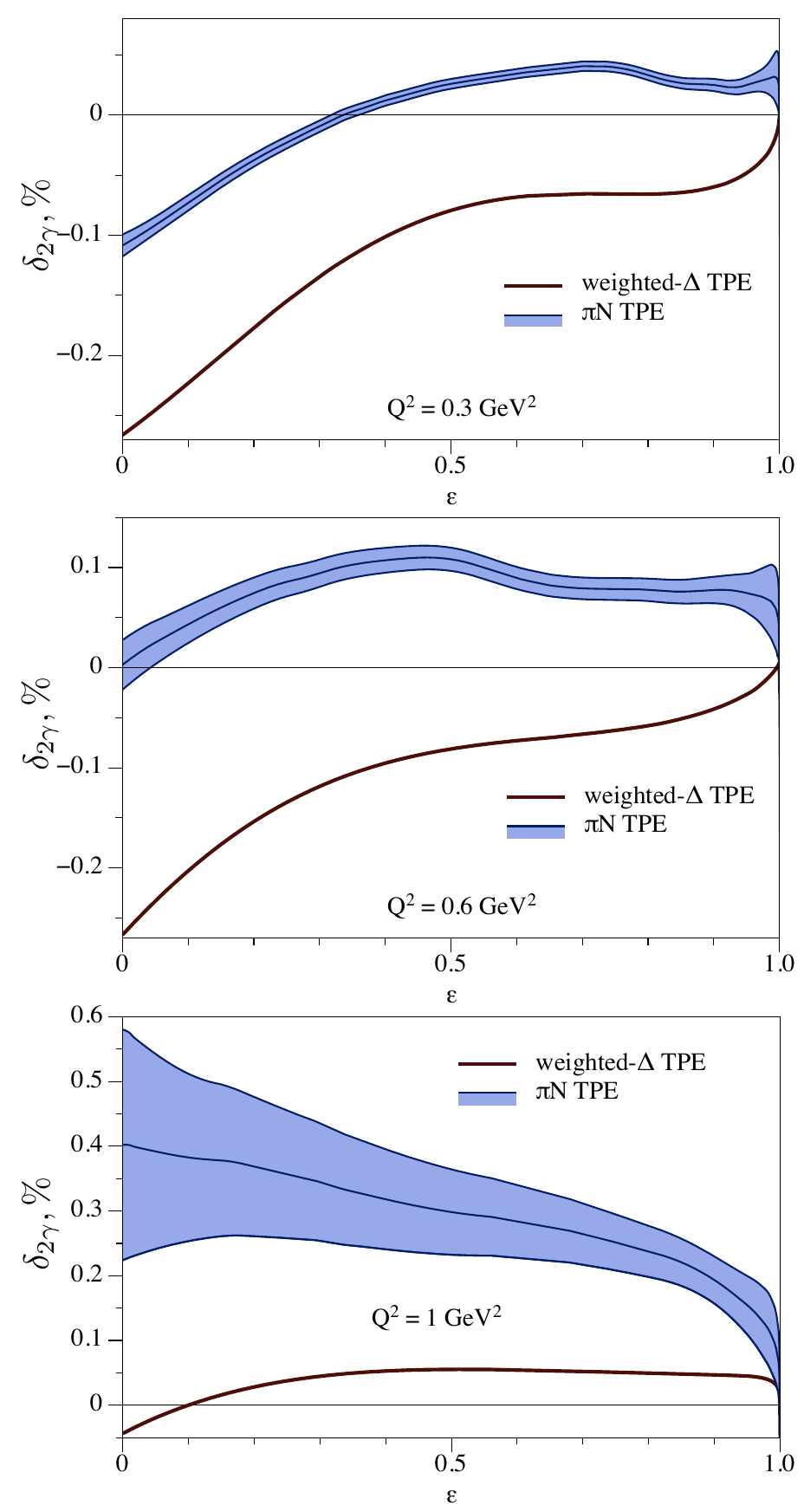}
\end{center}
\caption{The $\pi N$ intermediate state TPE correction to the unpolarized cross section for elastic electron-proton scattering, see Eq. (\ref{unpolarized_cross_section}), in comparison to the dispersive weighted-$\Delta$ model TPE result. For the $\pi N$ TPE result, the MAID fit is used up to $ W = 2~\mathrm{GeV}$, and the $ W > 2~\mathrm{GeV}$ behavior is approximated by Eqs. (\ref{f_1}), (\ref{f_2}). Upper plot: $Q^2 = 0.3~\mathrm{GeV}^2$, central plot: $Q^2 = 0.6~\mathrm{GeV}^2$, lower plot: $Q^2 = 1~\mathrm{GeV}^2$.}
\label{cross_section_reconstruction_piN}
\end{figure}

The resulting cross section corrections are shown in Fig. \ref{cross_section_reconstruction_piN} in comparison to the weighted-$\Delta$ model calculation. The amplitude uncertainties to the unpolarized cross section are added in quadrature. We see from Fig. \ref{cross_section_reconstruction_piN} that the $ \pi N $ TPE correction is always larger than the weighted-$\Delta$ model TPE. However, the $ \pi N$ contribution has a similar order of magnitude and shows an opposite sign at lower $Q^2$ and large $\varepsilon$.

\section{Results and discussion}
\label{sec5}

In this section, we provide a comparison of the dispersion relation calculation of the $ \pi N$ intermediate state contribution to TPE observables with recent experimental results \cite{Rachek:2014fam,Rimal:2016toz,Henderson:2016dea}. We also compare our results with previous TPE estimates of inelastic intermediate states in the near-forward approximation \cite{Tomalak:2015aoa}. The latter calculation provides an estimate of TPE corrections at low momentum transfer and small scattering angles through the unpolarized proton structure functions.

The TPE correction to the unpolarized cross section $\delta_{2 \gamma}$ can be directly accessed from the ratio of the positron-proton to electron-proton elastic scattering cross section $R_{2\gamma}$, in which it enters with different signs:
\ber \label{ratio_2gamma}
R_{2\gamma} = \frac{\sigma(e^{+} p)}{\sigma(e^{-} p)} = \frac{1 + \delta_{\mathrm{odd}} + \delta_{\mathrm{even}} -  \delta_{2 \gamma}}{1 - \delta_{\mathrm{odd}} + \delta_{\mathrm{even}} + \delta_{2 \gamma}} \approx  1 - 2 \delta_{2 \gamma}.
\eer
The approximation in the last step of Eq. (\ref{ratio_2gamma}) amounts to neglect the higher-order contributions of the charge-even radiative corrections $ \delta_{\mathrm{even}} $. Furthermore, we dropped the charge-odd radiative corrections $ \delta_{\mathrm{odd}} $, which are usually directly applied to the data.

In recent years several new measurements of the ratio of Eq. (\ref{ratio_2gamma}) were performed with a much improved precision in comparison to the early experiments from SLAC \cite{Mar:1968qd}. These new data come from the VEPP-3 storage ring in Novosibirsk \cite{Rachek:2014fam}, from the CLAS Collaboration at JLab \cite{Moteabbed:2013isu,Adikaram:2014ykv,Rimal:2016toz}, and from the OLYMPUS experiment at DESY \cite{Henderson:2016dea}.

\begin{figure}[H]
\begin{center}
\includegraphics[width=.8\textwidth]{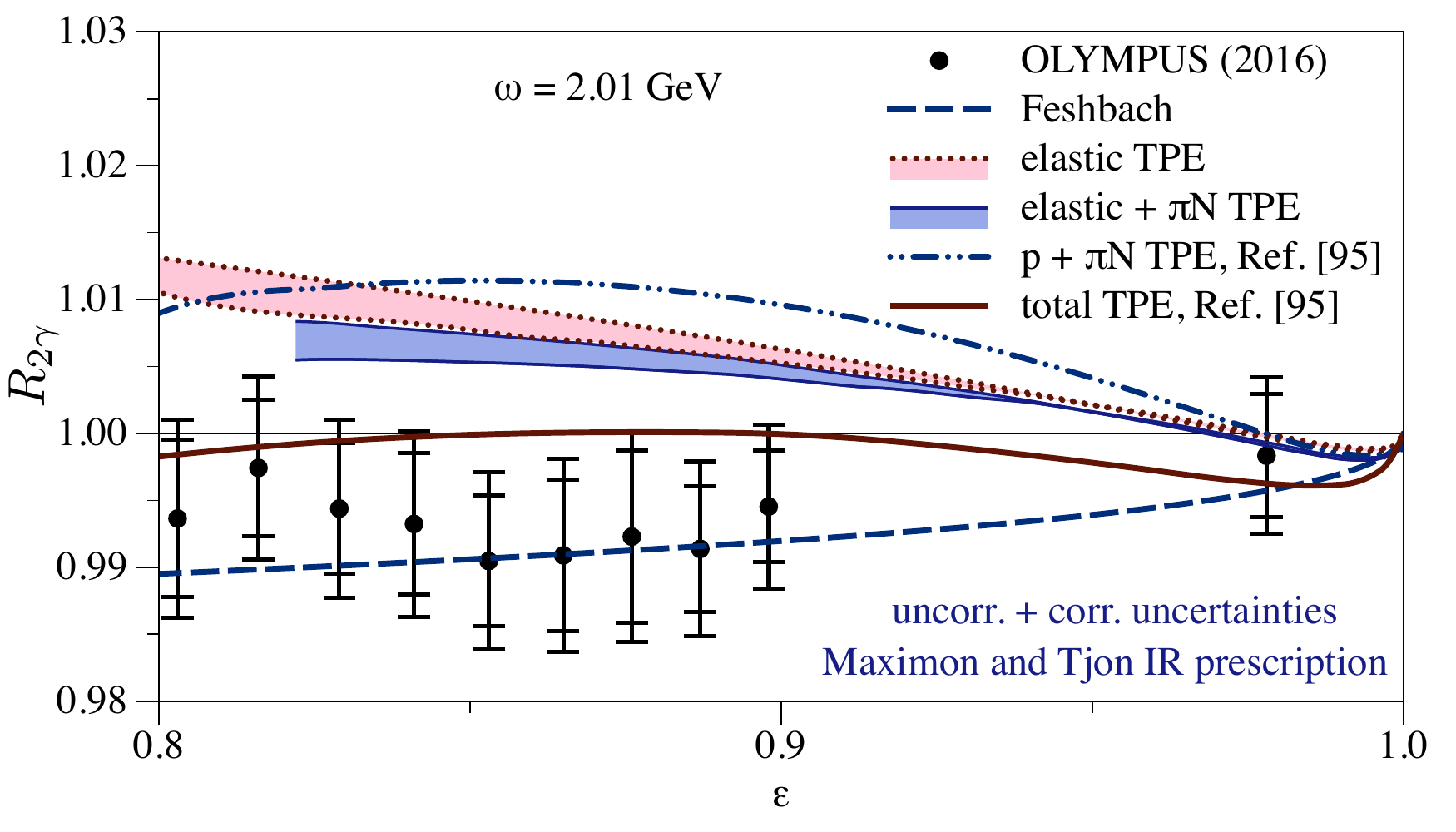}
\end{center}
\caption{The DR result for the elastic TPE and for the sum of the elastic and $\pi N$ TPE contributions to the $e^+ p$ over $e^- p$ elastic scattering cross section ratio $ R_{2 \gamma}$ for lepton beam energy $ \omega = 2.01~\mathrm{GeV}$, in comparison with the data from the Olympus Collaboration \cite{Henderson:2016dea}. We also show the Feshbach correction \cite{McKinley:1948zz}, as well as the total TPE and the sum of the proton + $\pi N$ contributions in the near-forward approximation  of Ref.~\cite{Tomalak:2015aoa}.}
\label{delta_OLYMPUS}
\begin{center}
\includegraphics[width=.8\textwidth]{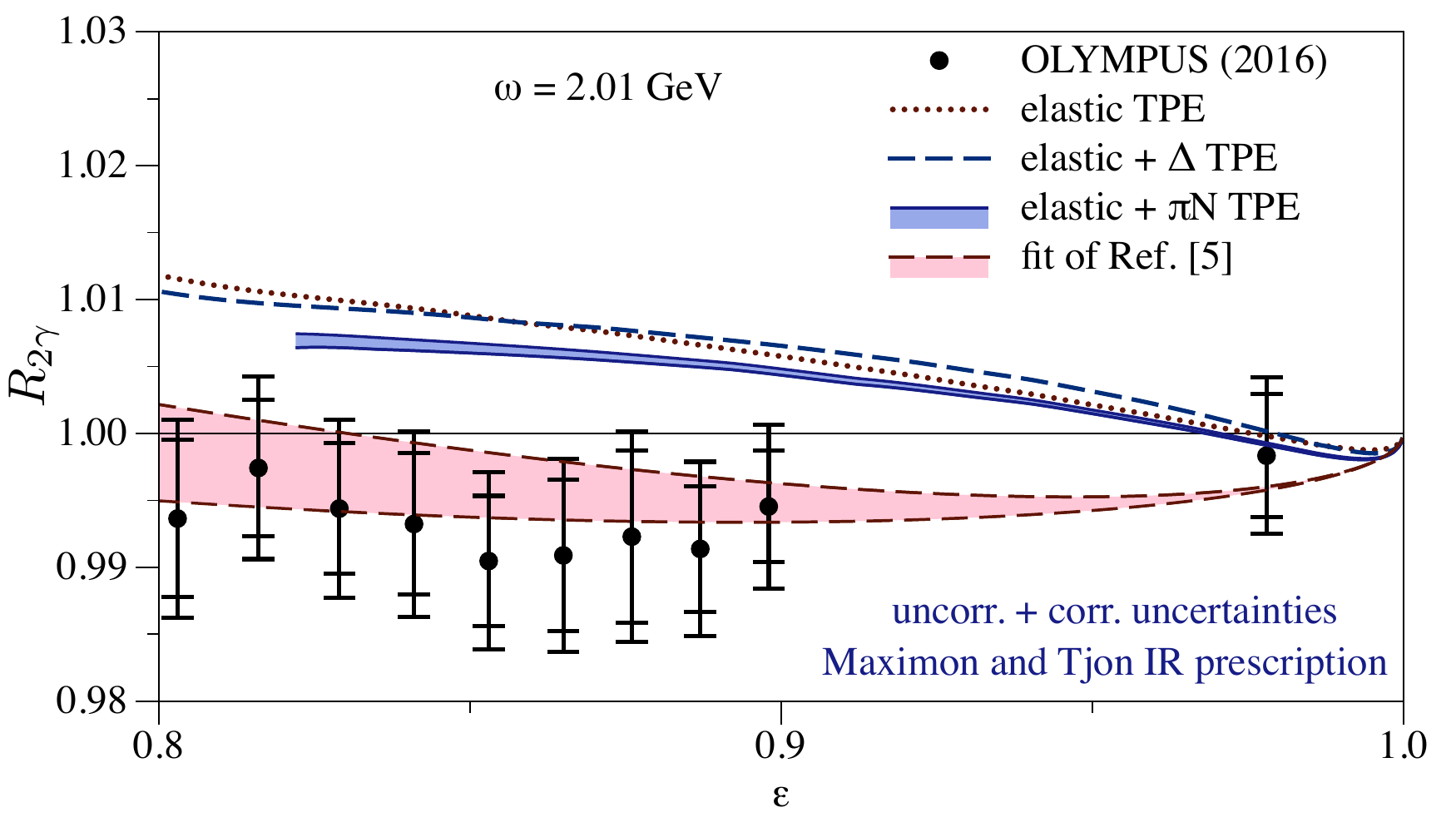}
\end{center}
\caption{The DR result for the elastic TPE and for the sum of the elastic + $\pi N$ TPE contributions to the $e^+ p$ over $e^- p$ elastic scattering cross section ratio $ R_{2 \gamma}$ in comparison with the sum of elastic + weighted-$\Delta$ calculation of Sec. \ref{sec3}, as well as with the phenomenological fit of Ref.~\cite{Bernauer:2013tpr}. The central value of the elastic contribution was used in this plot.}
\label{delta_OLYMPUS2}
\end{figure}

In Fig. \ref{delta_OLYMPUS}, we compare the dispersive evaluation of the $ \pi N$ intermediate state TPE contributions with the data of the OLYMPUS experiment \cite{Henderson:2016dea}, which measured the ratio $ R_{2 \gamma}$ using a 2.01 $\mathrm{GeV}$ lepton beam. We also show the Feshbach correction \cite{McKinley:1948zz} corresponding with the scattering on a heavy point charge, the elastic TPE, which includes the full nucleon electromagnetic structure, and the total TPE in the near-forward approximation of Ref.~\cite{Tomalak:2015aoa}. To study the relative contribution of other channels, we present the TPE correction in the near-forward approximation of Ref.~\cite{Tomalak:2015aoa} based on the comparison when using the total unpolarized proton structure functions as an input and when using its counterpart from the MAID2007 fit~\cite{Drechsel:1998hk,Drechsel:2007if}, which only includes the $\pi N$ channel. To evaluate the elastic TPE we exploit the FF fit to the unpolarized and polarization transfer world data \cite{Bernauer:2013tpr} and the analytical continuation method of Ref.~\cite{Tomalak:2014sva} for the central value. We estimate the $1 \sigma$ uncertainty bands of the elastic TPE by the difference when calculating the correction either with the empirical FFs or with a dipole form for the proton FFs. The IR divergences for all curves in Fig. \ref{delta_OLYMPUS} as well as for other plots in this section were subtracted according to the Maximon and Tjon prescription \cite{Maximon:2000hm}. We see from Fig. \ref{delta_OLYMPUS} that all theoretical curves are in agreement with the Feshbach correction in the forward limit $ \varepsilon \to 1$. Note that in the $ \varepsilon > 0.8$ region the OLYMPUS result is accidentally close to the Feshbach correction. As was mentioned in Ref.~\cite{Tomalak:2015aoa}, the proton form factor effect and the inelastic TPE contribution have different signs partially canceling each other. The dispersive result for the sum of elastic and $\pi N$ contributions is $~ 1\%$ above the experimental data at $Q^2 \gtrsim 0.624~\mathrm{GeV^2}$ ($\epsilon \lesssim 0.897 $) and is in agreement with the data point at the lowest momentum transfer as well as with the corresponding contribution in the near-forward approximation at large $\varepsilon$. The near-forward total TPE of Ref.~\cite{Tomalak:2015aoa}, which uses the forward proton structure functions as input to account for all inelastic intermediate states, describes the measurements surprisingly well even at relatively large momentum transfer beyond the expected region of applicability of such calculation. The comparison in Fig. \ref{delta_OLYMPUS} indicates that in the momentum transfer range $Q^2 \lesssim 1~\mathrm{GeV}^2$ ($\epsilon \gtrsim 0.809 $) the inelastic intermediate states reduce the TPE ratio $R_{2 \gamma}$ by around $1\%$-$1.5\%$, of which roughly half originates from $\pi N$ intermediate states and half from higher inelastic intermediate states.

In the following Fig. \ref{delta_OLYMPUS2}, we compare the dispersive evaluations of the sum of elastic + $ \pi N $ TPE with the sum of elastic + weighted-$\Delta$ TPE of Sec. \ref{sec3} and the phenomenological fit of Ref.~\cite{Bernauer:2013tpr}. The phenomenological fit of Ref.~\cite{Bernauer:2013tpr} provides a relatively good description of the experimental data. For the theoretical estimates, we first notice that all curves are in agreement with the Feshbach correction in the forward limit $ \varepsilon \to 1$. Recently, the narrow-$\Delta$ TPE correction was independently evaluated within a dispersion relation framework in Ref.~\cite{Blunden:2017nby}. Our result for the weighted-$\Delta$ TPE changes sign around $\varepsilon \approx 0.857$ in qualitative agreement with Ref.~\cite{Blunden:2017nby}. The account for the full $ \pi N$ intermediate state contribution moves the unsubtracted DR prediction closer to the data points in comparison to the $\Delta$ calculation. We may conclude that the account of higher intermediate states within the dispersive framework is necessary to improve the description of data for $R_{2 \gamma}$.

\begin{figure}[H]
\begin{center}
\includegraphics[width=1.\textwidth]{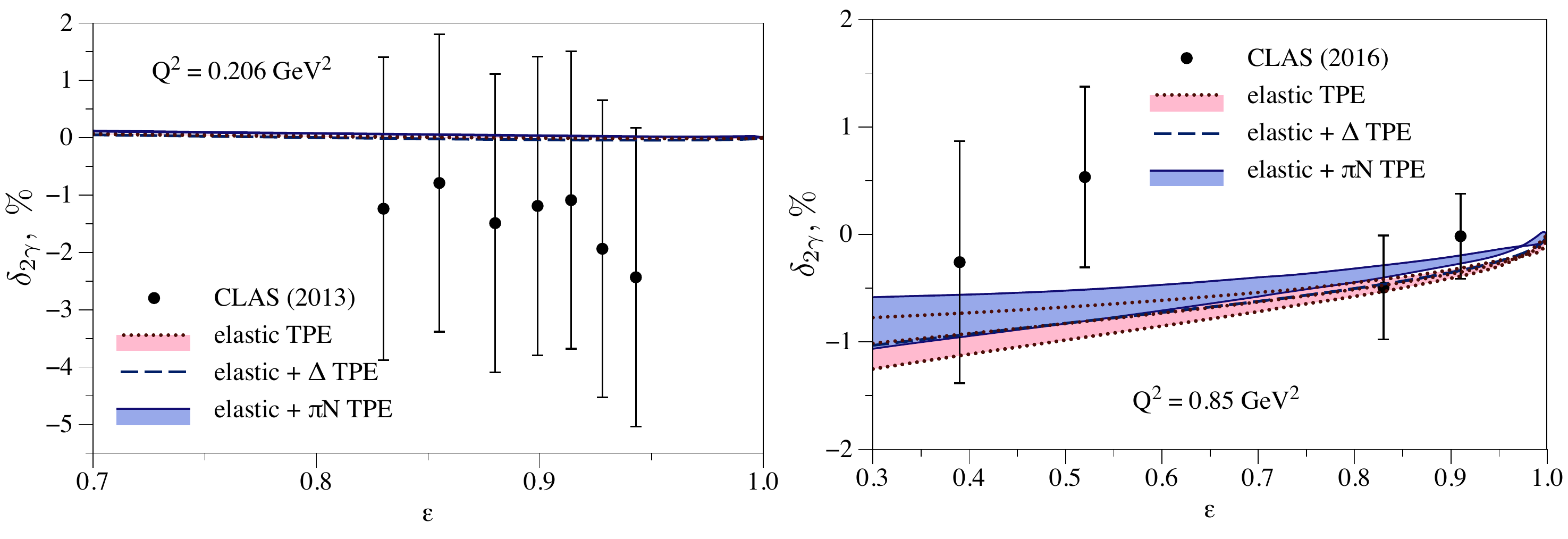}
\end{center}
\caption{Comparison of the unsubtracted DR prediction for the elastic, the weighted-$\Delta$ and $\pi N$ TPE correction for $ Q^2 = 0.206 ~ \mathrm{GeV}^2 $ (left panel) with the data of Ref.~\cite{Moteabbed:2013isu} and for $ Q^2 = 0.85 ~ \mathrm{GeV}^2 $ (right panel) with the data of Ref.~\cite{Rimal:2016toz}.}
\label{delta_CLAS_Zhan_larger}
\end{figure}

The CLAS Collaboration has performed measurements of $R_{2 \gamma}$ at relatively small values of the momentum transfer: $ Q^2 \approx 0.206 ~\mathrm{GeV}^2 $ \cite{Moteabbed:2013isu} and $ Q^2 = 0.85 ~ \mathrm{GeV}^2 $ \cite{Rimal:2016toz}. Neglecting the higher-order contributions of the charge-even radiative corrections and exploiting the total charge-even radiative correction factor from Ref.~\cite{Moteabbed:2013isu} $\delta_{\mathrm{even}} \approx -0.2$, the TPE contribution can be extracted as \cite{Moteabbed:2013isu}
\ber \label{CLAS_points}
\delta_{2 \gamma} \approx   \frac{1-R_{2 \gamma}}{2} (1 + \delta_{\mathrm{even}}).
\eer
In Fig. \ref{delta_CLAS_Zhan_larger}, we compare the elastic, the weighted-$\Delta$ and $ \pi N$ TPE corrections with the data from CLAS. 

The early CLAS measurements at $Q^2 \approx 0.206 ~\mathrm{GeV}^2 $ \cite{Moteabbed:2013isu} show large uncertainties which do not allow us to make strong conclusions. At this low $Q^2$ value, elastic, weighted-$\Delta$ and $\pi N$ TPE corrections are much smaller than $1~\%$.

The follow-up CLAS experiment of Ref. \cite{Rimal:2016toz} achieved a precision below the $1\%$ level as shown on Fig. \ref{delta_CLAS_Zhan_larger} (right panel) for $Q^2 = 0.85 ~ \mathrm{GeV}^2$. We notice that at $ Q^2 = 0.85 ~ \mathrm{GeV}^2 $, the account of  the $\pi N$ intermediate state contribution to TPE amplitudes on top of the elastic TPE improves the description of experimental data. Note, that the weighted-$\Delta$ TPE correction is much smaller than the $ \pi N$ TPE and changes sign in Fig. \ref{delta_CLAS_Zhan_larger} 

\begin{figure}[H]
\begin{center}
\includegraphics[width=0.86\textwidth]{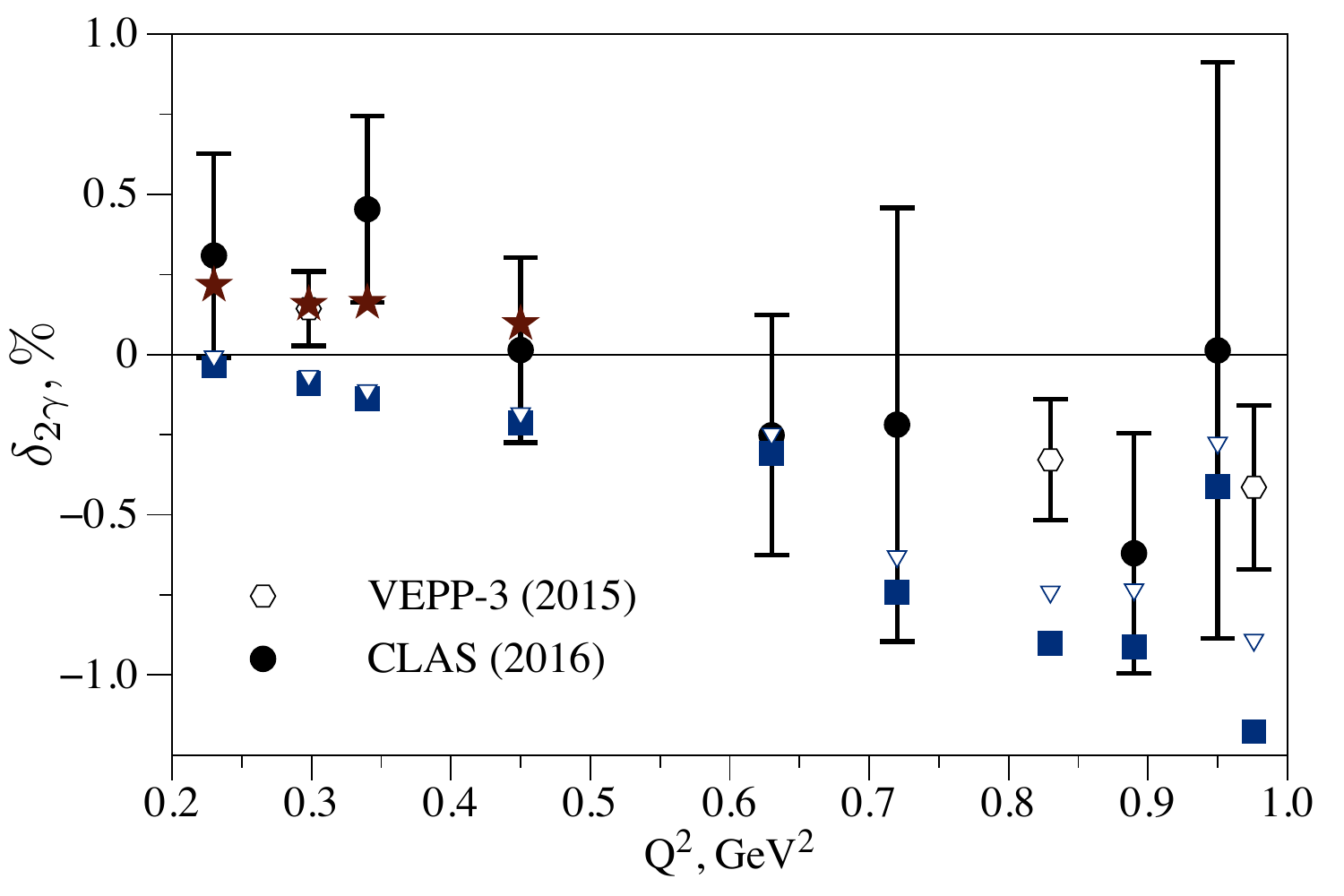}
\end{center}
\caption{TPE correction measurements of Refs.~\cite{Rachek:2014fam,Rimal:2016toz} in comparison with the elastic TPE (shown by squares), and the sum of elastic + $ \pi N$ TPE (shown by hollow triangles). For $Q^2 < 0.5~\mathrm{GeV}^2$, we also show the comparison with the total near-forward TPE of Ref.~\cite{Tomalak:2015aoa} (shown by stars). The CLAS \cite{Rimal:2016toz} data points correspond to the kinematics of Table \ref{table_CLAS}. The VEPP-3 \cite{Rachek:2014fam} data points correspond to  $Q^2 = 0.298~\mathrm{GeV}^2,~\varepsilon = 0.93$; $Q^2 = 0.83~\mathrm{GeV}^2,~\varepsilon = 0.4$; and $Q^2 = 0.976~\mathrm{GeV}^2,~\varepsilon = 0.27$. The VEPP-3 data points were renormalized according to the empirical fit of Ref.~\cite{Bernauer:2013tpr} by a procedure which is explained in Ref. \cite{Rachek:2014fam}.}
\label{delta_CLAS_Zhan_088}
\end{figure}

In Fig. \ref{delta_CLAS_Zhan_088}, we compare the $Q^2$ dependence of $\delta_{2 \gamma}$ for VEPP-3 data \cite{Rachek:2014fam} and CLAS data \cite{Rimal:2016toz} with the elastic TPE (central value), the sum of elastic + $ \pi N$ TPE (central value), and total TPE in the near-forward approximation. We provide the kinematics of the CLAS data points~\cite{Rimal:2016toz} in Table \ref{table_CLAS}. The CLAS values of $\delta_{2 \gamma}$ were obtained using Eq. (\ref{CLAS_points}).
\begin{table}[H] 
\begin{center}
\begin{tabular}{|c|c|c|c|c|c|c|c|}
\noalign{\vskip 5mm}    
\hline
$ \varepsilon $ & $ ~ 0.92~$ & $ ~0.89 ~$ & $  ~0.89 ~$  &  $ ~ 0.89 ~$ & $ ~0.45~ $ & $~ 0.45~ $ & $ ~0.88 ~$   \\ \hline
$ Q^2,~\mathrm{GeV}^2 $ & $ 0.23 $ & $ 0.34 $ & $ 0.45 $  &  $ 0.63 $ & $ 0.72 $ & $ 0.89 $ & $ 0.89 $  \\ \hline
\end{tabular}
\caption{Kinematics of the CLAS experiment of Ref. \cite{Rimal:2016toz}.}  \label{table_CLAS}
\end{center}
\end{table}
We notice from Fig. \ref{delta_CLAS_Zhan_088} that the CLAS data points are in agreement with the total TPE correction in the near-forward approximation. However, the VEPP-3 data point of Ref.~\cite{Rachek:2014fam} at $ Q^2 = 0.298~\mathrm{GeV}^2$ agrees with the total TPE only after the renormalization procedure as it is described in Ref. \cite{Rachek:2014fam}.We perform the renormalization in Fig. \ref{delta_CLAS_Zhan_088} according to the empirical fit of Ref.~\cite{Bernauer:2013tpr}. Accounting for the $\pi N$ intermediate state within the dispersive framework, the data are described better than by the elastic contribution solely. However, the CLAS data point$~Q^2 = 0.34~\mathrm{GeV}^2,~\varepsilon = 0.89$ and all VEPP-3 data points differ from the dispersion relation result by more than $ 1 \sigma$. An additional correction of the same sign as the inelastic $\pi N$ contribution is needed to reconcile the difference between theory and these data points. Multiparticle states TPE contribution can at least partially reconcile this discrepancy.

\begin{figure}[H]
\begin{center}
\includegraphics[width=.68\textwidth]{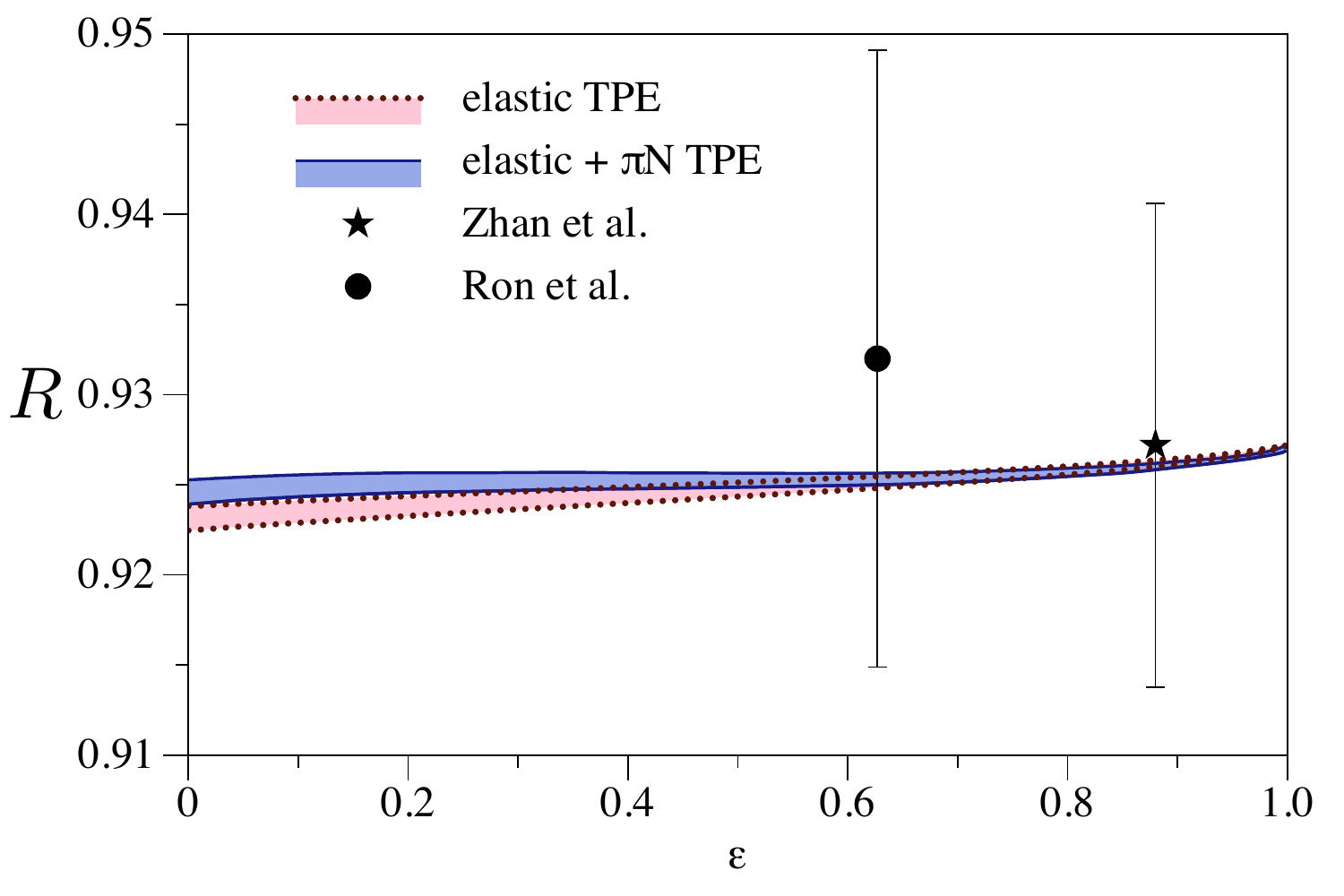}
\end{center}
\caption{Comparison of the unsubtracted DR prediction for the elastic and $ \pi N $ TPE correction for the ratio $ R = -\mu_p \sqrt{\frac{1+\varepsilon}{\varepsilon}\tau_P} \frac{P_t}{P_l}$  at $ Q^2 = 0.3 ~ \mathrm{GeV}^2 $ with the data of Refs.~\cite{Zhan:2011ji,Ron:2011rd}.}
\label{ptpl_CLAS_Zhan}
\end{figure}

Finally, the ratio $ P_t / P_l $, Eq. (\ref{polarization_observables1}), was measured at the low momentum transfer region for $ Q^2 = 0.298 ~\mathrm{GeV}^2 $ \cite{Zhan:2011ji} and $ Q^2 = 0.308 ~\mathrm{GeV}^2 $ \cite{Ron:2011rd} in Hall A at JLab. In absence of TPE corrections, the $ R = -\mu_p \sqrt{\frac{1+\varepsilon}{\varepsilon}\tau_P} \frac{P_t}{P_l} $ ratio at fixed $Q^2$ is $ \varepsilon$ independent. We compare the polarization transfer data points to the elastic TPE and the sum of the elastic + $ \pi N$ TPE in Fig. \ref{ptpl_CLAS_Zhan}. The calculation of TPE amplitudes was performed at $ Q^2 = 0.3 ~\mathrm{GeV}^2$ exploiting the proton elastic FFs of Ref.~\cite{Zhan:2011ji} in Eq. (\ref{polarization_observables1}). The account for the $\pi N$ intermediate state just slightly modifies the unsubtracted DR prediction for the elastic contribution  in the region of available data, confirming the small value of TPE correction to the polarization transfer observables.

\section{Conclusions and Outlook}
\label{sec6}

In this work we have accounted for the pion-nucleon ($\pi N$) TPE correction within a dispersion relation framework extending the kinematical coverage of Ref.~\cite{Tomalak:2016vbf} to the momentum transfers $ 0.064~\mathrm{GeV}^2 \lesssim Q^2 \lesssim 1~\mathrm{GeV}^2$. On the example of a box graph calculation with a $\Delta$-intermediate state we have developed and successfully tested a new method for the analytical continuation of the invariant amplitudes into the unphysical region, which relies on the knowledge of the imaginary parts in the physical region solely. Generalizing the method to the $ \pi N$ intermediate state, we evaluated the $\pi N$ TPE contribution using the MAID 2007 parametrization for the pion electroproduction amplitudes as input and estimated uncertainties of our method. We have made a comparison of the dispersion relation results with recent measurements of the TPE correction to the unpolarized elastic electron-proton scattering cross section \cite{Rachek:2014fam,Rimal:2016toz,Henderson:2016dea}. With account of the $\pi N$ intermediate state, the TPE correction comes closer to the experimental data in comparison with the elastic contribution only confirming the cancellation between the inelastic TPE and the proton form factor effects, which was previously found in Ref.~\cite{Tomalak:2015aoa}. An additional correction of the order of $ 1~\%$  is needed to describe the OLYMPUS and VEPP-3 data points within the error bars. A near-forward calculation in terms of inclusive proton structure functions indicates that multiparticle intermediate states, especially $\pi \pi N$, can be responsible for this difference. However, the evaluated $\pi N$ TPE correction can be now exploited for a precise extraction of the proton magnetic radius and the proton magnetic form factor at low values of $Q^2$.
 
\section*{Acknowledgments}
We thank Lothar Tiator for useful discussions and providing us with MAID programs, Dalibor Djukanovic for providing us with the access to computer resources, Volodymyr Schubny for discussions about the narrow-$\Delta$ contribution. We acknowledge the computing time granted on the supercomputer Mogon at Johannes Gutenberg University Mainz (hpc.uni-mainz.de). This work was supported by the Deutsche Forschungsgemeinschaft DFG in part through the Collaborative Research Center [The Low-Energy Frontier of the Standard Model (SFB 1044)], and in part through the Cluster of Excellence [Precision Physics, Fundamental Interactions and Structure of Matter (PRISMA)].


\newpage
\begin{thebibliography}{99}

\bibitem{Mcallister:1956ng} 
  R.~W.~Mcallister and R.~Hofstadter,
  Phys.\ Rev.\  {\bf 102}, 851 (1956).


\bibitem{Hofstadter:1956qs} 
  R.~Hofstadter,
  Rev.\ Mod.\ Phys.\  {\bf 28}, 214 (1956).


\bibitem{Rosenbluth:1950yq} 
  M.~N.~Rosenbluth,
  Phys.\ Rev.\  {\bf 79}, 615 (1950).


\bibitem{Bernauer:2010wm} 
  J.~C.~Bernauer {\it et al.} [A1 Collaboration],
  Phys.\ Rev.\ Lett.\  {\bf 105}, 242001 (2010).


\bibitem{Bernauer:2013tpr} 
  J.~C.~Bernauer {\it et al.} [A1 Collaboration],
  Phys.\ Rev.\ C {\bf 90}, no. 1, 015206 (2014).

\bibitem{Lorenz:2012tm} 
  I.~T.~Lorenz, H.-W.~Hammer and U.~G.~Meissner,
  Eur.\ Phys.\ J.\ A {\bf 48}, 151 (2012).
  
\bibitem{Lorenz:2014vha} 
  I.~T.~Lorenz and U.~G.~Meissner,
  Phys.\ Lett.\ B {\bf 737}, 57 (2014).

\bibitem{Lorenz:2014yda} 
  I.~T.~Lorenz, U.~G.~Meißner, H.-W.~Hammer and Y.-B.~Dong,
  Phys.\ Rev.\ D {\bf 91}, no. 1, 014023 (2015).


\bibitem{Lee:2015jqa} 
  G.~Lee, J.~R.~Arrington and R.~J.~Hill,
  Phys.\ Rev.\ D {\bf 92}, no. 1, 013013 (2015).


\bibitem{Arrington:2015ria} 
  J.~Arrington and I.~Sick,
  J.\ Phys.\ Chem.\ Ref.\ Data {\bf 44}, 031204 (2015).


\bibitem{Arrington:2015yxa} 
  J.~Arrington,
  J.\ Phys.\ Chem.\ Ref.\ Data {\bf 44}, 031203 (2015).


\bibitem{Griffioen:2015hta} 
  K.~Griffioen, C.~Carlson and S.~Maddox,
  Phys.\ Rev.\ C {\bf 93}, no. 6, 065207 (2016).


\bibitem{Higinbotham:2015rja} 
  D.~W.~Higinbotham, A.~A.~Kabir, V.~Lin, D.~Meekins, B.~Norum and B.~Sawatzky,
  Phys.\ Rev.\ C {\bf 93}, no. 5, 055207 (2016).



 
\bibitem{Mo:1968cg} 
  L.~W.~Mo and Y.~S.~Tsai,
  Rev.\ Mod.\ Phys.\  {\bf 41}, 205 (1969).
  
\bibitem{Maximon:2000hm} 
  L.~C.~Maximon and J.~A.~Tjon,
  Phys.\ Rev.\ C {\bf 62}, 054320 (2000).
 
\bibitem{Hill:2016gdf} 
  R.~J.~Hill,
  Phys.\ Rev.\ D {\bf 95}, no. 1, 013001 (2017).


\bibitem{Pohl:2010zza} 
  R.~Pohl {\it et al.},
  Nature {\bf 466}, 213 (2010).


\bibitem{Antognini:1900ns} 
  A.~Antognini {\it et al.},
  Science {\bf 339}, 417 (2013).


\bibitem{Mohr:2012tt} 
  P.~J.~Mohr, B.~N.~Taylor and D.~B.~Newell,
  Rev.\ Mod.\ Phys.\  {\bf 84}, 1527 (2012).


\bibitem{Carlson:2015jba} 
  C.~E.~Carlson,
  Prog.\ Part.\ Nucl.\ Phys.\  {\bf 82}, 59 (2015).


\bibitem{Hill:2017wzi} 
  R.~J.~Hill,
  EPJ Web Conf.\  {\bf 137}, 01023 (2017).


\bibitem{Pachucki:1996zza} 
  K.~Pachucki,
  Phys.\ Rev.\ A {\bf 53}, 2092 (1996).


\bibitem{Faustov:1999ga} 
  R.~N.~Faustov and A.~P.~Martynenko,
  Phys.\ Atom.\ Nucl.\  {\bf 63}, 845 (2000)
  [Yad.\ Fiz.\  {\bf 63}, 915 (2000)].


\bibitem{Pineda:2002as} 
  A.~Pineda,
  Phys.\ Rev.\ C {\bf 67}, 025201 (2003).


\bibitem{Pineda:2004mx} 
  A.~Pineda,
  Phys.\ Rev.\ C {\bf 71}, 065205 (2005).


\bibitem{Nevado:2007dd} 
  D.~Nevado and A.~Pineda,
  Phys.\ Rev.\ C {\bf 77}, 035202 (2008).


\bibitem{Carlson:2011zd} 
  C.~E.~Carlson and M.~Vanderhaeghen,
  Phys.\ Rev.\ A {\bf 84}, 020102 (2011).


\bibitem{Hill:2012rh} 
  R.~J.~Hill, G.~Lee, G.~Paz and M.~P.~Solon,
  Phys.\ Rev.\ D {\bf 87}, 053017 (2013).


\bibitem{Birse:2012eb} 
  M.~C.~Birse and J.~A.~McGovern,
  Eur.\ Phys.\ J.\ A {\bf 48}, 120 (2012).


\bibitem{Alarcon:2013cba} 
  J.~M.~Alarcon, V.~Lensky and V.~Pascalutsa,
  Eur.\ Phys.\ J.\ C {\bf 74}, no. 4, 2852 (2014).


\bibitem{Gorchtein:2013yga} 
  M.~Gorchtein, F.~J.~Llanes-Estrada and A.~P.~Szczepaniak,
  Phys.\ Rev.\ A {\bf 87}, no. 5, 052501 (2013).


\bibitem{Peset:2014jxa} 
  C.~Peset and A.~Pineda,
  Nucl.\ Phys.\ B {\bf 887}, 69 (2014).


\bibitem{Tomalak:2015hva} 
  O.~Tomalak and M.~Vanderhaeghen,
  Eur.\ Phys.\ J.\ C {\bf 76}, no. 3, 125 (2016).


\bibitem{Caprini:2016wvy} 
  I.~Caprini,
  Phys.\ Rev.\ D {\bf 93}, no. 7, 076002 (2016).


\bibitem{Hill:2016bjv} 
  R.~J.~Hill and G.~Paz,
  Phys.\ Rev.\ D {\bf 95}, no. 9, 094017 (2017).


\bibitem{Pohl:2016tqq} 
  R.~Pohl [CREMA Collaboration],
  J.\ Phys.\ Soc.\ Jap.\  {\bf 85}, no. 9, 091003 (2016).


\bibitem{Dupays:2003zz} 
  A.~Dupays, A.~Beswick, B.~Lepetit, C.~Rizzo and D.~Bakalov,
  Phys.\ Rev.\ A {\bf 68}, 052503 (2003).


\bibitem{Adamczak:2016pdb} 
  A.~Adamczak {\it et al.} [FAMU Collaboration],
  JINST {\bf 11}, no. 05, P05007 (2016).


\bibitem{Ma:2016etb} 
  Y.~Ma {\it et al.},
  Int.\ J.\ Mod.\ Phys.\ Conf.\ Ser.\  {\bf 40}, 1660046 (2016).


\bibitem{Zemach:1956zz} 
  A.~C.~Zemach,
  Phys.\ Rev.\  {\bf 104}, 1771 (1956).


\bibitem{Iddings:1959zz} 
  C.~K.~Iddings and P.~M.~Platzman,
  Phys.\ Rev.\  {\bf 113}, 192 (1959).


\bibitem{Iddings:1965zz} 
  C.~K.~Iddings,
  Phys.\ Rev.\  {\bf 138}, B446 (1965).


\bibitem{Drell:1966kk} 
  S.~D.~Drell and J.~D.~Sullivan,
  Phys.\ Rev.\  {\bf 154}, 1477 (1967).


\bibitem{Faustov:1966} 
	R. ~N. ~Faustov,
    Nucl.\ Phys. {\bf 75}, 669 (1966).

\bibitem{Faustov:1970} 
	G. ~M. ~Zinovjev, B. ~V. ~Struminski, R. ~N. ~Faustov, and V. ~L. ~Chernyak, 
    Sov.\ J.\ Nucl.\ Phys. {\bf 11}, 715 (1970).

\bibitem{Bodwin:1987mj} 
  G.~T.~Bodwin and D.~R.~Yennie,
  Phys.\ Rev.\ D {\bf 37}, 498 (1988).


\bibitem{Faustov:2001pn} 
  R.~N.~Faustov, E.~V.~Cherednikova and A.~P.~Martynenko,
  Nucl.\ Phys.\ A {\bf 703}, 365 (2002).


\bibitem{Carlson:2008ke} 
  C.~E.~Carlson, V.~Nazaryan and K.~Griffioen,
  Phys.\ Rev.\ A {\bf 78}, 022517 (2008).


\bibitem{Carlson:2011af} 
  C.~E.~Carlson, V.~Nazaryan and K.~Griffioen,
  Phys.\ Rev.\ A {\bf 83}, 042509 (2011).


\bibitem{Hagelstein:2015egb} 
  F.~Hagelstein, R.~Miskimen and V.~Pascalutsa,
  Prog.\ Part.\ Nucl.\ Phys.\  {\bf 88}, 29 (2016).


\bibitem{Peset:2016wjq} 
  C.~Peset and A.~Pineda,
  JHEP {\bf 1704}, 060 (2017).

\bibitem{Tomalak:2017owk} 
  O.~Tomalak,
  Eur.\ Phys.\ J.\ C {\bf 77}, no. 8, 517 (2017).
  
\bibitem{Tomalak:2017npu} 
  O.~Tomalak,
  arXiv:1708.02509 [hep-ph].

\bibitem{Denig:2016dqo} 
  A.~Denig,
  AIP Conf.\ Proc.\  {\bf 1735}, 020006 (2016).

\bibitem{Akhiezer:1968ek} 
  A.~I.~Akhiezer and M.~P.~Rekalo,
  Sov.\ Phys.\ Dokl.\  {\bf 13}, 572 (1968)
  [Dokl.\ Akad.\ Nauk Ser.\ Fiz.\  {\bf 180}, 1081 (1968)].


\bibitem{Akhiezer:1974em} 
  A.~I.~Akhiezer and M.~P.~Rekalo,
  Sov.\ J.\ Part.\ Nucl.\  {\bf 4}, 277 (1974)
  [Fiz.\ Elem.\ Chast.\ Atom.\ Yadra {\bf 4}, 662 (1973)].


\bibitem{Dombey:1969wk} 
  N.~Dombey,
  Rev.\ Mod.\ Phys.\  {\bf 41}, 236 (1969).


\bibitem{Dombey:1969wi} 
  N.~Dombey,
  Phys.\ Lett.\  {\bf 29B}, 588 (1969).


\bibitem{Punjabi:2015bba} 
  V.~Punjabi, C.~F.~Perdrisat, M.~K.~Jones, E.~J.~Brash and C.~E.~Carlson,
  Eur.\ Phys.\ J.\ A {\bf 51}, 79 (2015).


\bibitem{Jones:1999rz} 
  M.~K.~Jones {\it et al.} [Jefferson Lab Hall A Collaboration],
  Phys.\ Rev.\ Lett.\  {\bf 84}, 1398 (2000).


\bibitem{Gayou:2001qd} 
  O.~Gayou {\it et al.} [Jefferson Lab Hall A Collaboration],
  Phys.\ Rev.\ Lett.\  {\bf 88}, 092301 (2002).


\bibitem{Punjabi:2005wq} 
  V.~Punjabi {\it et al.},
  Phys.\ Rev.\ C {\bf 71}, 055202 (2005)
  Erratum: [Phys.\ Rev.\ C {\bf 71}, 069902 (2005)].


\bibitem{Puckett:2010ac} 
  A.~J.~R.~Puckett {\it et al.},
  Phys.\ Rev.\ Lett.\  {\bf 104}, 242301 (2010).


\bibitem{Guichon:2003qm} 
  P.~A.~M.~Guichon and M.~Vanderhaeghen,
  Phys.\ Rev.\ Lett.\  {\bf 91}, 142303 (2003).


\bibitem{Blunden:2003sp} 
  P.~G.~Blunden, W.~Melnitchouk and J.~A.~Tjon,
  Phys.\ Rev.\ Lett.\  {\bf 91}, 142304 (2003).


\bibitem{Afanasev:2002gr} 
  A.~Afanasev, I.~Akushevich and N.~P.~Merenkov,
  hep-ph/0208260.


\bibitem{Gorchtein:2004ac} 
  M.~Gorchtein, P.~A.~M.~Guichon and M.~Vanderhaeghen,
  Nucl.\ Phys.\ A {\bf 741}, 234 (2004).


\bibitem{Pasquini:2004pv} 
  B.~Pasquini and M.~Vanderhaeghen,
  Phys.\ Rev.\ C {\bf 70}, 045206 (2004).


\bibitem{Chen:2004tw} 
  Y.~C.~Chen, A.~Afanasev, S.~J.~Brodsky, C.~E.~Carlson and M.~Vanderhaeghen,
  Phys.\ Rev.\ Lett.\  {\bf 93}, 122301 (2004).


\bibitem{Afanasev:2005mp} 
  A.~V.~Afanasev, S.~J.~Brodsky, C.~E.~Carlson, Y.~C.~Chen and M.~Vanderhaeghen,
  Phys.\ Rev.\ D {\bf 72}, 013008 (2005).


\bibitem{Borisyuk:2008db} 
  D.~Borisyuk and A.~Kobushkin,
  Phys.\ Rev.\ D {\bf 79}, 034001 (2009).


\bibitem{Kivel:2009eg} 
  N.~Kivel and M.~Vanderhaeghen,
  Phys.\ Rev.\ Lett.\  {\bf 103}, 092004 (2009).


\bibitem{Kivel:2012vs} 
  N.~Kivel and M.~Vanderhaeghen,
  JHEP {\bf 1304}, 029 (2013).


\bibitem{Wells:2000rx} 
  S.~P.~Wells {\it et al.} [SAMPLE Collaboration],
  Phys.\ Rev.\ C {\bf 63}, 064001 (2001).


\bibitem{Maas:2004pd} 
  F.~E.~Maas {\it et al.},
  Phys.\ Rev.\ Lett.\  {\bf 94}, 082001 (2005).


\bibitem{Meziane:2010xc} 
  M.~Meziane {\it et al.} [GEp2gamma Collaboration],
  Phys.\ Rev.\ Lett.\  {\bf 106}, 132501 (2011).


\bibitem{Guttmann:2010au} 
  J.~Guttmann, N.~Kivel, M.~Meziane and M.~Vanderhaeghen,
  Eur.\ Phys.\ J.\ A {\bf 47}, 77 (2011).


\bibitem{BalaguerRios:2012uk} 
  D.~Balaguer Rios,
  Nuovo Cim.\ C {\bf 035N04}, 198 (2012).


\bibitem{Abrahamyan:2012cg} 
  S.~Abrahamyan {\it et al.} [HAPPEX and PREX Collaborations],
  Phys.\ Rev.\ Lett.\  {\bf 109}, 192501 (2012).


\bibitem{Waidyawansa:2013yva} 
  B.~P.~Waidyawansa [Qweak Collaboration],
  AIP Conf.\ Proc.\  {\bf 1560}, 583 (2013).


\bibitem{Kumar:2013yoa} 
  K.~S.~Kumar, S.~Mantry, W.~J.~Marciano and P.~A.~Souder,
  Ann.\ Rev.\ Nucl.\ Part.\ Sci.\  {\bf 63}, 237 (2013).


\bibitem{Nuruzzaman:2015vba} 
  Nuruzzaman [Qweak Collaboration],
  arXiv:1510.00449 [nucl-ex].


\bibitem{Zhang:2015kna} 
  Y.~W.~Zhang {\it et al.},
  Phys.\ Rev.\ Lett.\  {\bf 115}, no. 17, 172502 (2015).


\bibitem{Carlson:2007sp} 
  C.~E.~Carlson and M.~Vanderhaeghen,
  Ann.\ Rev.\ Nucl.\ Part.\ Sci.\  {\bf 57}, 171 (2007).


\bibitem{Arrington:2011dn} 
  J.~Arrington, P.~G.~Blunden and W.~Melnitchouk,
  Prog.\ Part.\ Nucl.\ Phys.\  {\bf 66}, 782 (2011).


\bibitem{Rachek:2014fam} 
  I.~A.~Rachek {\it et al.},
  Phys.\ Rev.\ Lett.\  {\bf 114}, no. 6, 062005 (2015).


\bibitem{Moteabbed:2013isu} 
  M.~Moteabbed {\it et al.} [CLAS Collaboration],
  Phys.\ Rev.\ C {\bf 88}, 025210 (2013).

\bibitem{Adikaram:2014ykv} 
  D.~Adikaram {\it et al.} [CLAS Collaboration],
  Phys.\ Rev.\ Lett.\  {\bf 114}, 062003 (2015).


\bibitem{Rimal:2016toz} 
  D.~Rimal {\it et al.} [CLAS Collaboration],
  Phys.\ Rev.\ C {\bf 95}, no. 6, 065201 (2017).


\bibitem{Henderson:2016dea} 
  B.~S.~Henderson {\it et al.} [OLYMPUS Collaboration],
  Phys.\ Rev.\ Lett.\  {\bf 118}, no. 9, 092501 (2017).


\bibitem{Afanasev:2017gsk} 
  A.~Afanasev, P.~G.~Blunden, D.~Hasell and B.~A.~Raue,
  Prog.\ Part.\ Nucl.\ Phys.\  {\bf 95}, 245 (2017).


\bibitem{McKinley:1948zz} 
  W.~A.~McKinley and H.~Feshbach,
  Phys.\ Rev.\  {\bf 74}, 1759 (1948).


\bibitem{Brown:1970te} 
  R.~W.~Brown,
  Phys.\ Rev.\ D {\bf 1}, 1432 (1970).


\bibitem{Gorchtein:2014hla} 
  M.~Gorchtein,
  Phys.\ Rev.\ C {\bf 90}, no. 5, 052201 (2014).


\bibitem{Tomalak:2015aoa} 
  O.~Tomalak and M.~Vanderhaeghen,
  Phys.\ Rev.\ D {\bf 93}, no. 1, 013023 (2016).


\bibitem{Kondratyuk:2005kk} 
  S.~Kondratyuk, P.~G.~Blunden, W.~Melnitchouk and J.~A.~Tjon,
  Phys.\ Rev.\ Lett.\  {\bf 95}, 172503 (2005).


\bibitem{Graczyk:2013pca} 
  K.~M.~Graczyk,
  Phys.\ Rev.\ C {\bf 88}, 065205 (2013).


\bibitem{Zhou:2014xka} 
  H.~Q.~Zhou and S.~N.~Yang,
  Eur.\ Phys.\ J.\ A {\bf 51}, no. 8, 105 (2015).


\bibitem{Kondratyuk:2007hc} 
  S.~Kondratyuk and P.~G.~Blunden,
  Phys.\ Rev.\ C {\bf 75}, 038201 (2007).


\bibitem{Blunden:2017nby} 
  P.~G.~Blunden and W.~Melnitchouk,
  Phys.\ Rev.\ C {\bf 95}, no. 6, 065209 (2017).


\bibitem{Gorchtein:2006mq} 
  M.~Gorchtein,
  Phys.\ Lett.\ B {\bf 644}, 322 (2007).


\bibitem{Borisyuk:2008es} 
  D.~Borisyuk and A.~Kobushkin,
  Phys.\ Rev.\ C {\bf 78}, 025208 (2008).


\bibitem{Borisyuk:2012he} 
  D.~Borisyuk and A.~Kobushkin,
  Phys.\ Rev.\ C {\bf 86}, 055204 (2012).


\bibitem{Borisyuk:2013hja} 
  D.~Borisyuk and A.~Kobushkin,
  Phys.\ Rev.\ C {\bf 89}, no. 2, 025204 (2014).


\bibitem{Borisyuk:2015xma} 
  D.~Borisyuk and A.~Kobushkin,
  Phys.\ Rev.\ C {\bf 92}, no. 3, 035204 (2015).


\bibitem{Tomalak:2014sva} 
  O.~Tomalak and M.~Vanderhaeghen,
  Eur.\ Phys.\ J.\ A {\bf 51}, no. 2, 24 (2015).


\bibitem{Tomalak:2016vbf} 
  O.~Tomalak, B.~Pasquini and M.~Vanderhaeghen,
  Phys.\ Rev.\ D {\bf 95}, no. 9, 096001 (2017).


\bibitem{Drechsel:1998hk} 
  D.~Drechsel, O.~Hanstein, S.~S.~Kamalov and L.~Tiator,
  Nucl.\ Phys.\ A {\bf 645}, 145 (1999).


\bibitem{Drechsel:2007if} 
  D.~Drechsel, S.~S.~Kamalov and L.~Tiator,
  Eur.\ Phys.\ J.\ A {\bf 34}, 69 (2007).

  
\bibitem{Jones:1972ky} 
  H.~F.~Jones and M.~D.~Scadron,
  Annals Phys.\  {\bf 81}, 1 (1973).


\bibitem{Pascalutsa:2007wz} 
  V.~Pascalutsa and M.~Vanderhaeghen,
  Phys.\ Rev.\ D {\bf 76}, 111501 (2007).


\bibitem{Warren:2003ma} 
  G.~Warren {\it et al.} [Jefferson Lab E93-026 Collaboration],
  Phys.\ Rev.\ Lett.\  {\bf 92}, 042301 (2004).


\bibitem{Hahn:2000jm} 
  T.~Hahn,
  Nucl.\ Phys.\ Proc.\ Suppl.\  {\bf 89}, 231 (2000).


\bibitem{vanOldenborgh:1989wn} 
  G.~J.~van Oldenborgh and J.~A.~M.~Vermaseren,
  Z.\ Phys.\ C {\bf 46}, 425 (1990).


\bibitem{Froissart:1961ux} 
  M.~Froissart,
  Phys.\ Rev.\  {\bf 123}, 1053 (1961).


\bibitem{Tomalak_PhD}
	O. Tomalak, Ph.D. dissertation, Johannes Gutenberg-Universit\"at Mainz, 2016.

\bibitem{Mar:1968qd} 
  J.~Mar {\it et al.},
  Phys.\ Rev.\ Lett.\  {\bf 21}, 482 (1968).


\bibitem{Zhan:2011ji} 
  X.~Zhan {\it et al.},
  Phys.\ Lett.\ B {\bf 705}, 59 (2011).


\bibitem{Ron:2011rd} 
  G.~Ron {\it et al.} [Jefferson Lab Hall A Collaboration],
  Phys.\ Rev.\ C {\bf 84}, 055204 (2011).

\end{thebibliography}
\end{document}